\documentclass[sigconf]{acmart}
\usepackage{subcaption}
\usepackage{pifont}
\usepackage{enumitem}
\usepackage{xcolor}
\usepackage{makecell}
\usepackage{algorithm}
\usepackage{graphicx}
\usepackage{multirow}

\definecolor{myblue}{RGB}{79,173,234}

\AtBeginDocument{%
  }

\copyrightyear{2026}
\acmYear{2026}
\setcopyright{cc}
\setcctype{by}
\acmConference[CHI '26]{Proceedings of the 2026 CHI Conference on Human Factors in Computing Systems}{April 13--17, 2026}{Barcelona, Spain}
\acmBooktitle{Proceedings of the 2026 CHI Conference on Human Factors in Computing Systems (CHI '26), April 13--17, 2026, Barcelona, Spain}
\acmPrice{}
\acmDOI{10.1145/3772318.3790445}
\acmISBN{979-8-4007-2278-3/2026/04}

\newcommand{\cmark}{\textcolor{green!60!black}{\ding{51}}}  
\newcommand{\xmark}{\textcolor{red}{\ding{55}}}             
\newcommand{\red}[1]{\textcolor{black}{#1}}
\newcommand{\rr}[1]{\textcolor{black}{#1}}

\newcommand{\sysname}{\textit{LubDubDecoder}}

\newcommand{\squishlist}{\begin{itemize}[itemsep=1pt,parsep=2pt,topsep=3pt,partopsep=0pt,leftmargin=0em, itemindent=1em,labelwidth=1em,labelsep=0.5em]}
\newcommand{\squishend}{\end{itemize}}
\newcommand{\squishenum}{%
  \begin{enumerate}[
label=\textbf{(\arabic*)},itemsep=1pt,parsep=2pt,topsep=3pt,partopsep=0pt,
    leftmargin=0pt,
    labelindent=0pt,
    itemindent=0pt,
    labelsep=0.5em,
    listparindent=0pt
  ]}

\newcommand{\squishsubenum}{%
  \begin{enumerate}[
    label=\textbf{(\alph*)},
    itemsep=1pt,parsep=2pt,topsep=0pt,partopsep=0pt,
    leftmargin=0pt,
    labelindent=0pt,
    itemindent=0pt,
    labelsep=0.5em,
    listparindent=0pt
  ]}

\newcommand{\squishenumend}{\end{enumerate}}

\newcommand{\ssquishlist}{%
  \begin{itemize}[
    label={},
    itemsep=1pt,parsep=2pt,topsep=3pt,partopsep=0pt,
    leftmargin=0pt,
    labelindent=0pt,
    itemindent=0pt,
    labelsep=0.5em,
    listparindent=0pt
  ]}

\newcommand{\ssquishsublist}{%
  \begin{itemize}[
    label={},
    itemsep=1pt,parsep=2pt,topsep=0pt,partopsep=0pt,
    leftmargin=0pt,
    labelindent=0pt,
    itemindent=0pt,
    labelsep=0.5em,
    listparindent=0pt
  ]}

\newcommand{\ssquishend}{\end{itemize}}

\begin{document}

\title{{\sysname}: Bringing Micro-Mechanical \\Cardiac Monitoring to Hearables}

\begin{abstract}
We present {\sysname}, a system that enables fine-grained monitoring of micro-cardiac vibrations associated with the opening and closing of heart valves across a range of hearables. Our system transforms the built-in speaker, the only transducer common to all hearables, into an acoustic sensor that captures the coarse ``lub-dub'' heart sounds, leverages their shared temporal and spectral structure to reconstruct the subtle seismocardiography (SCG) and gyrocardiography (GCG) waveforms, and extract the timing of key micro-cardiac events. In an IRB-approved feasibility study with \red{25} users, our system achieves correlations of 0.88--0.95 compared to chest-mounted reference measurements in within-user and cross-user evaluations, and generalizes to unseen hearables using a zero-effort adaptation scheme with a correlation of 0.91. Our system is robust across remounting sessions and music playback. 
\end{abstract}



\author{Siqi Zhang}
\authornote{Co-primary authors}
\affiliation{Carnegie Mellon University \country{USA}}
\email{siqiz2@andrew.cmu.edu}

\author{Xiyuxing Zhang}
\authornotemark[1]
\affiliation{Tsinghua University\country{China}}
\email{zxyx22@mails.tsinghua.edu.cn}

\author{Duc Vu}
\authornotemark[1]
\affiliation{Michigan State University\country{USA}}
\email{vuduc2@msu.edu}

\author{Tao Qiang}
\authornotemark[1]
\affiliation{Shanghai Jiao Tong University\country{China}}
\email{riderdecade@sjtu.edu.cn}

\author{Clara Palacios}
\affiliation{Comillas Pontifical University\country{Spain}}
\email{clara.palacios.spain@gmail.com}

\author{Jiangyifei Zhu}
\affiliation{Carnegie Mellon University \country{USA}}
\email{jiangyiz@andrew.cmu.edu}

\author{Yuntao Wang}
\authornote{Co-corresponding authors}
\affiliation{Tsinghua University\country{China}}
\email{yuntaowang@tsinghua.edu.cn}

\author{Mayank Goel}
\affiliation{Carnegie Mellon University \country{USA}}
\email{mayankgoel@cmu.edu}

\author{Justin Chan}
\authornotemark[2]
\affiliation{Carnegie Mellon University \country{USA}}
\email{justinchan@cmu.edu}

\renewcommand{\shortauthors}{Zhang et al.}

\begin{teaserfigure}
\centering
\includegraphics[keepaspectratio, width=.9\linewidth]{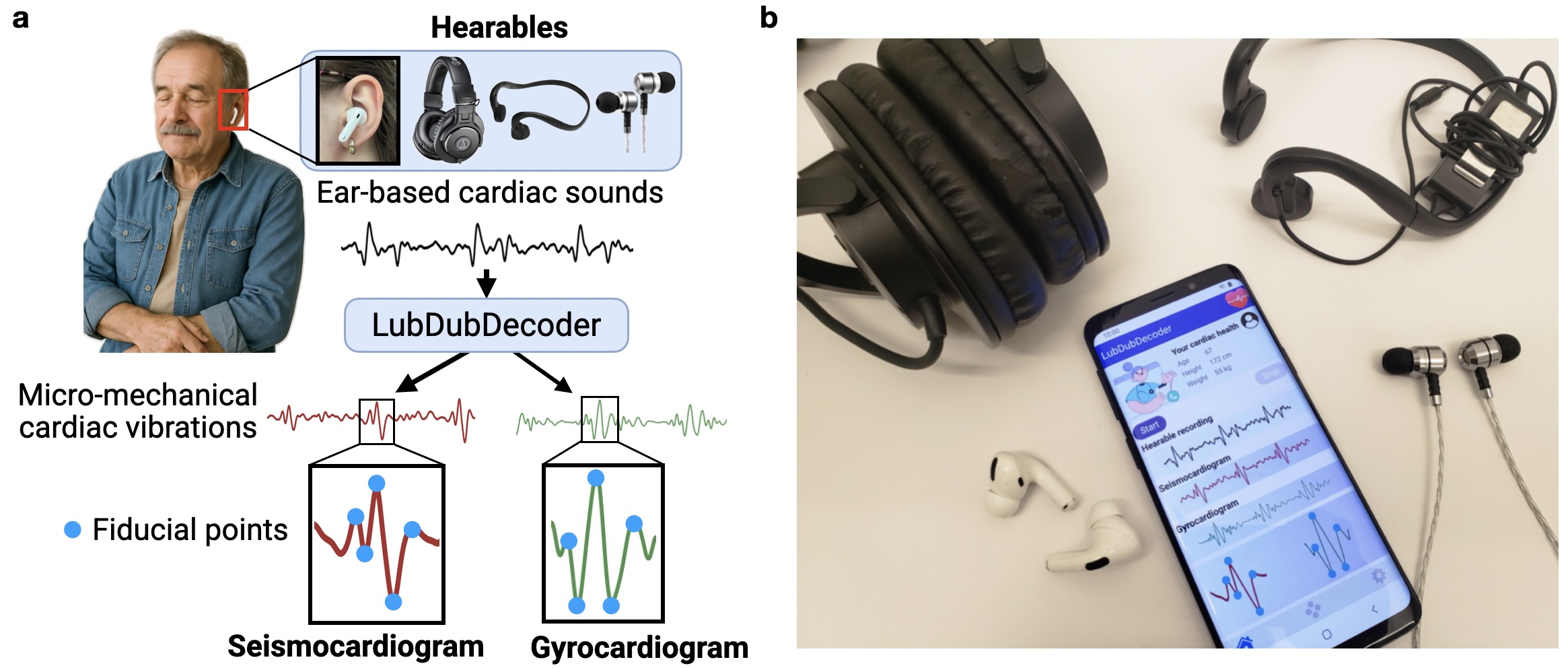}
  \vspace{-1em}
  \caption{{\bf \sysname System overview. (a)} {\sysname} reconstructs the heart's micro-mechanical signals from coarse-grained heart sounds recorded at the ear using microphones and speakers across a range of hearables. It identifies the timing of key micro-cardiac events, and enables hands-free monitoring of cardiovascular health in everyday scenarios. {\bf (b)} Smartphone app user interface with hearables.}
  \label{fig:fig1}
  \Description{This figure illustrates the LubDubDecoder system. It provides a detailed system overview as a large diagram. On the top left, an image of a person wearing an earbud represents the input source, with other compatible "Hearables" shown above, including over-ear headphones, bone-conduction headphones, wireleass earbuds and wired earphones. Arrows lead from this to a central diagram labeled "LubDubDecoder," which processes ear-based cardiac sounds to reconstruct micro-mechanical cardiac vibrations. Below this, two graphs show the reconstructed signals: a Seismocardiogram (SCG) and a Gyrocardiogram (GCG), with key "Fiducial points" marked as blue dots on their waveforms.}
\end{teaserfigure}

\maketitle

\section{Introduction}

Fine-grained micro-mechanical vibrations of the chest reveal the timing of key cardiac events, such as the opening and closing of heart valves, and can be used for screening heart conditions including heart failure, ischemia, and atrial fibrillation~\cite{shandhi2019seismocardiography,pankaala2016detection,korzeniowska2005usefulness,inan2018novel}. These vibrations are measured using seismocardiography (SCG) and gyrocardiography (GCG), which capture chest motion through linear acceleration and rotation, respectively~\cite{taebi2019recent,zanetti2013seismocardiography,jafari2017gyrocardiography,siecinski2020gyrocardiography}.

Collecting these signals, however, typically requires a clinical setting in which the patient lies down, removes their shirt, and is instrumented with accelerometers and gyroscopes~\cite{zanetti2013seismocardiography}. Furthermore, recordings are typically limited to a few minutes due to time constraints in busy clinics, as well as patient discomfort during prolonged sessions. 

The ability to enable unobtrusive monitoring of these micro-cardiac events outside the clinic allows for measurements in a naturalistic setting free from clinical stressors such as the ``white coat effect'' that can spuriously elevate cardiovascular measurements~\cite{franklin2013white}. Such monitoring can support timely interventions including lifestyle changes to physical activity and diet~\cite{verdecchia1995white,pickering2002white,guimaraes2010effects,shibata2018effect,boutouyrie2011pharmacological}. It is also particularly useful for populations with mobility restrictions, such as older adults who are at highest risk of cardiac abnormalities~\cite{andrawes2005prevention,piccini2014clinical}.

Given cardiac events send a pressure wave through the body that causes body parts to vibrate, the use of smartphone inertial measurement units (IMUs) appear at first to be an accessible way to capture these signals. However, the measured waveforms are highly sensitive to minute variations in placement: small shifts of just a few centimeters on the chest can substantially alter the recordings due to differences in local tissue composition and device coupling, making it difficult to compare results across sessions~\cite{sandler2020documenting}. This challenge is amplified when measurements are performed by lay users outside clinical settings, where consistent and precise placement can be difficult to ensure.

Hearables, by contrast, occupy a fixed anatomical position in the ear, minimizing placement variability across removal and re-mounting sessions. Given that micro-cardiac vibrations propagate from the chest to the ear canal, hearables present themselves as a promising platform for convenient everyday monitoring of these signals.

We present {\sysname}, the first hearable system that bring micro-mechanical cardiac monitoring out of the clinic and enables monitoring in everyday scenarios across a range hearables (Fig.~\ref{fig:fig1}). While prior work~\cite{fu2025enabling} has reconstructed SCG using IMU-equipped hearables, IMUs are typically only found on true wireless stereo earbuds~\cite{fu2025enabling} or higher-end earphones such as those supporting active noise cancellation~\cite{chen2024exploring}. Other approaches measure SCG using IMUs on smartphones~\cite{wang2018seismo,lee2021discrimination,hossein2024smartphone} and contactless mmWave radar~\cite{ha2020contactless}, however, these require strapping or holding the phone against the chests, or constraining users to remain in a fixed location and orientation. In contrast, our approach is broadly compatible across hearables \rr{including low-cost earphones} and enables monitoring across environments, including during music playback.


However, there are several challenges in building such a system. 

\squishenum
\item {\bf Measuring micro-cardiac events at the ear.} Cardiac micro-vibrations are conventionally measured using chest-mounted IMUs because they are mechanical vibrations. However, IMUs are not available on all hearables. Further, these mechanical vibrations are attenuated and distorted as they propagate through the body channel consisting of bone, muscle, fat, and skin~\cite{chen2024exploring}, making them difficult to sense at the ear. 

Our key insight is that the familiar ``lub-dub'' sounds of the heart beating can be measured from the ear~\cite{butkow2023heart,chen2024exploring}, and leveraged to reconstruct the fine-grained SCG and GCG waveforms, enabling accurate extraction of the timing of micro-cardiac events. The intuition here is that because the heart sounds and vibratory SCG and GCG signals both originate from the same heart pumping mechanics, they carry shared temporal and spectral information~\cite{voyatzoglou2022introduction} which can be learned to derive a cross-modal mapping between the signals. 

To achieve this, we transform the speaker, \textit{the only transducer common to all hearables}, into a microphone to capture and amplify the soft, coarse-grained heart sounds from the ear and reconstruct the SCG and GCG signal. Our system works across hearables with different designs \rr{and varied price points} including bone-conduction earphones that contact the outer ear, over-ear headphones that enclose the ears, as well as both wired and wireless earbuds.

\item {\bf Adapting to individual physiology.} As individuals differ in physiology, the channel between the heart and ear varies from person to person. To account for these effects, our system requires only a brief calibration lasting a few seconds, where the user places their smartphone against the chest while recording simultaneously from the hearable and the phone’s IMU. This lightweight human-in-the-loop process is used to finetune a pretrained model to preserve the timing of micro-cardiac events. 

\textit{We note that periodic calibration is standard practice across both clinical and consumer cardiac monitoring due to drift in physiological signals from changes in body state (e.g. BMI) and skin properties. Clinic-grade cardiac monitoring devices~\cite{jones2003measuring}, and commercial smart devices such as the CE-marked Aktiia wristband~\cite{vybornova2021blood} and Samsung Galaxy Watch~\cite{swatch} require monthly recalibration to maintain accuracy. Our approach follows this established industry practice.}

\item {\bf Generalizing across hearables.} Measured heart sounds at the ear will vary across hearables due to sensor variability. As a result physiological signal mappings learned from one device cannot automatically translate to another device. To generalize to new hearable devices not seen during training, we apply a zero-effort normalization strategy that does not require any explicit calibration effort by the user. Specifically, ear-based cardiac sounds from the new device are normalized against recordings from a previously seen device, which serves as a reference to compute a set of weights. These weights are then applied to all subsequent recordings from the new device, enabling zero-effort adaptation.
\squishenumend

\noindent \red{This work contributes to the HCI community through:}
\squishlist
\item \red{A novel sensing approach that repurposes the in-ear speakers for cardiac sensing, transforming everyday hearables into an equitable health monitoring platform that remains accessible to individuals with limited income or low technology literacy.}
\item \red{An automated labeling pipeline for high-throughput segmentation and annotation of cardiac signals that can be applied to other cardiac health monitoring contexts.}
\item \red{Experimental protocols and a motion artifact detection algorithm for assessing feasibility in free living deployments across everyday static and dynamic activities and during music playback, which other researchers can build on.}
\squishend

\noindent Below we summarize the main technical contributions of our system:
\squishlist
\item {\sysname} reconstructs the fine-grained SCG and GCG waveforms from ear-based heart sounds and extracts the timing of key micro-cardiac events in an IRB-approved feasibility study with \red{25} subjects, achieving Pearson correlations between 0.88--0.95 compared to conventional chest-mounted measurements in both within-user and cross-user evaluations.
\item Our approach generalizes across different hearable designs in hardware by leveraging built-in speakers and microphones, and in software through a zero-effort adaptation scheme that achieves waveform correlations of 0.91 in cross-device testing.
\item Our system maintains robust performance across remounting sessions without recalibration, and in the presence of music playback from the hearable.
\item Our motion artifact removal pipeline detects and removes unwanted motion events with 97.7 $\pm$ 2.2\% accuracy.
\item We perform a user experience survey demonstrating high levels of ease of use \red{(4.6 of 5)} and system trustworthiness \red{(4.3 of 5)}.
\squishend

\noindent \rr{{\bf Contribution to hearables and cardiac sensing research in HCI.}} \rr{Conceptually, to the best of our knowledge, we are the first HCI system that is focused around using the {\it built-in speaker}, the only transducer common to all hearables, to perform cardiac monitoring across broader range of hearables of varying price points. In contrast, prior works use dedicated IMUs or in-ear microphones which are not present on most classes of hearables in particular low-cost earphones~\cite{fu2025enabling}. {\it We show that despite weaker acoustic output, speaker-based sensing can enable robust cardiac monitoring at comparable performance to in-ear mics (Fig.~\ref{fig:cross_device_main}).} This specific design choice is what allows us to have the benefits of greater ubiquity without trading off system performance.}

\rr{We focus on extracting SCG/GCG signals as it is unique amongst cardiac signals in its ability to measure {\it sub-audible mechanical events not measurable by PCG} and can assess cardiac conditions~\cite{iftikhar2018multiclass,pandia2012extracting,pankaala2016detection,sahoo2018design,salerno1991seismocardiography,solar2017classification,tadi2016real,taebi2017grouping,tavakolian2010characterization,tavakolian2010estimating,zakeri2016analyzing}, and in some cases even outperforming ECG electrical signals~\cite{wilson1993diagnostic}. This particular set of cardiac signals is under-explored in comparison to prior works which have focused on extracting ECG~\cite{butkow2023heart}, PCG~\cite{chen2024exploring}, respiration volume monitoring~\cite{liu2025earmeter}, and arterial pressure waveform~\cite{christofferson2025artearial}.}

\rr{Enabling practical SCG/GCG monitoring requires contributions from a system design perspective:}
\squishenum
\item \rr{{\it Cross condition characterization.} We perform a detailed characterization of micro-cardiac waveform variability in {\bf Sec.~\ref{sec:variability}} for intra-session, inter-session, inter-user and inter-device scenarios which motivates the design of our system, and reveals {\bf (a)} the susceptibility of conventional smartphone-based SCG/GCG monitoring to placement variability {\bf (b)} the comparative stability of in-ear SCG/GCG signals even after remounting {\bf (c)} the need for per-person and per-device calibration.}
\item \rr{{\it Novel calibration framework.} Based on these findings, we introduce in {\bf Sec.~\ref{sec:calib}} {\bf (i)} a human-in-the-loop adaptation procedure for new individuals which incorporates a brief simultaneous measurement using the smartphone on the chest and the hearable, and {\bf (ii)} a zero-effort normalization strategy for new hearables without any explicit effort on the part of the user that normalizes the frequency response of a new device with recordings from a previously seen device.}
\item \rr{{\it Micro-cardiac event timing reconstruction.} With this distinct setup design and formulation, we are able to reconstruct the timing of fiducial points from the ear in cross-session, cross-user and cross-device scenarios ({\bf Sec.~\ref{sec:timing}}) which is not shown in prior works~\cite{fu2025enabling}.}
\squishenumend

\noindent \rr{{\bf Ethics and consent.} This work includes {\bf (1)} a feasibility study evaluating the cardiac waveform reconstruction performance of our system, {\bf (2)} a post study user experience survey, and {\bf (3)} recordings to characterize the cardiac signals and train a motion artifact detection classifier. For each of these components, informed written consent was obtained prior to participation and prior to any data collection. All these procedures were conducted under IRB approval (STUDY2025\_00000155) and complied with relevant ethical regulations.}

\section{Background and related work}

\begin{figure}[t]
\centering
\includegraphics[keepaspectratio, width=\linewidth]
    {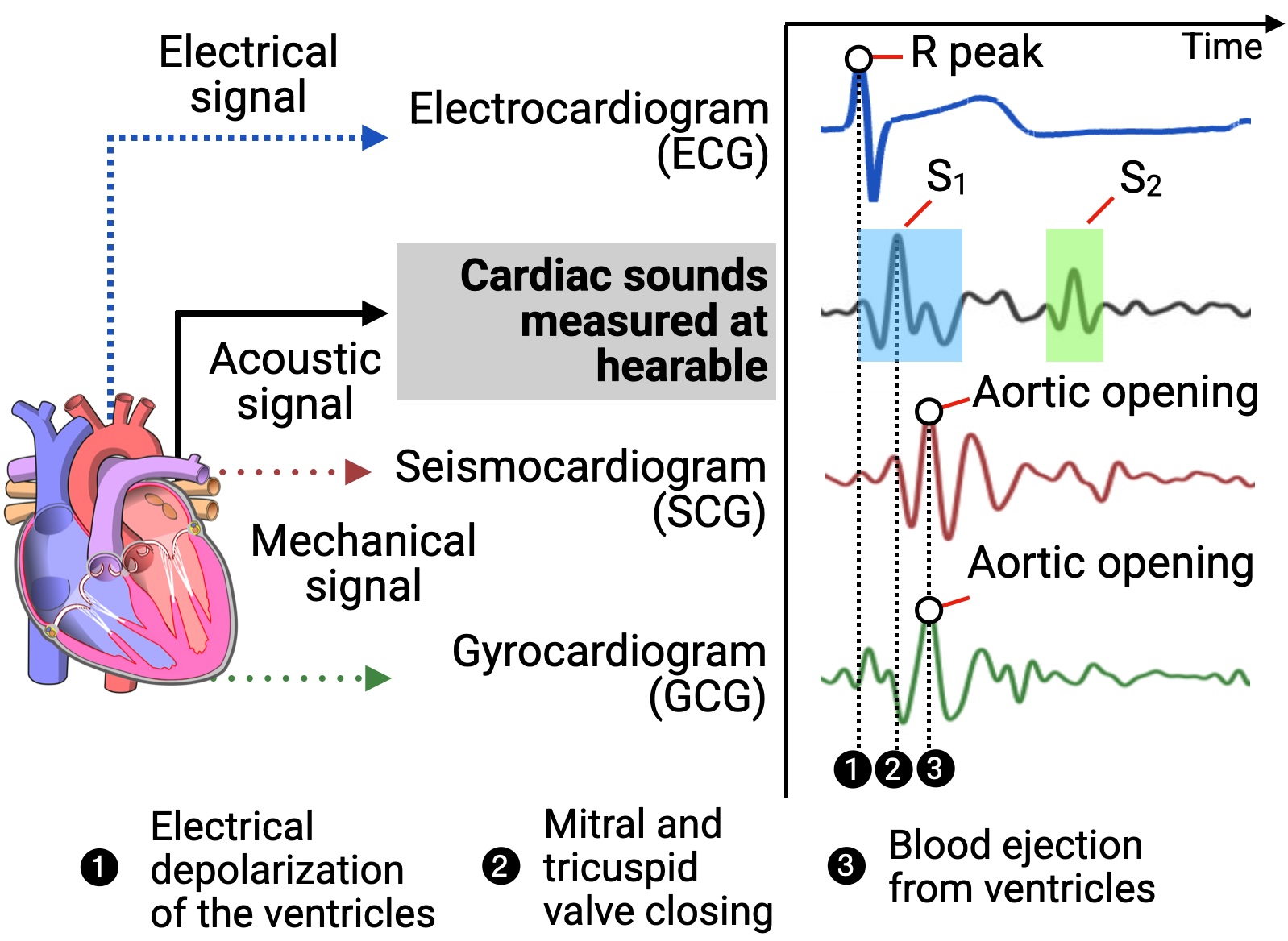}
    \vspace{-1em}
  \caption{{\bf Cardiac signal timing diagram.} Each heartbeat begins with the electrical depolarization of the ventricles, measurable by the ECG. This is followed by the mitral valve closing which creates the ``lub'' (S1) sound of the heart, which generates an acoustic signal detectable at the ear. Later in the cardiac cycle, the aortic valve closes, producing the ``dub'' (S2) sound. The heart’s mechanical motion can be captured at using SCG and GCG.} 
  \label{fig:physiology}
    \Description{This figure illustrates the synchronized relationship between the electrical, acoustic, and mechanical signals of the heart, with a focus on their temporal alignment. The left side of the figure displays a simplified anatomical drawing of the human heart, showing three distinct signal types originating from its activity: an electrical signal leading to an Electrocardiogram (ECG), an acoustic signal corresponding to Cardiac sounds as measured by a hearable device, and a mechanical signal associated with both a Seismocardiogram (SCG) and a Gyrocardiogram (GCG). The right side of the figure presents a time-aligned visualization of these four signals on a shared timeline. The top signal is the ECG, which shows a typical waveform with a prominent R peak. A vertical dashed line connects the R peak to a point marked as 1, which corresponds to the electrical depolarization of the ventricles. Below the ECG, the cardiac sound signal waveform displays two distinct sounds, S1 and S2, which are highlighted by a light blue box and a light green box, respectively. A second vertical dashed line from S1 connects to a point marked as 2, indicating the mitral and tricuspid valve closing. The bottom two signals are the SCG and GCG, which show a prominent event labeled Aortic opening. A third vertical dashed line connects this event to a point marked as 3, which signifies blood ejection from the ventricles.}
\end{figure}

\begin{table*}[t]
\centering
\small
\setlength{\tabcolsep}{3pt}
\renewcommand{\arraystretch}{1.1}

\resizebox{\textwidth}{!}{%
\begin{tabular}{|p{2.6cm}|p{2.4cm}|p{3.1cm}|p{2.1cm}|p{3.1cm}|p{2.6cm}|}
\hline
\textbf{Reference} &
\multicolumn{1}{c|}{\textbf{Sensor}} &
\multicolumn{1}{c|}{\textbf{Micro-mechanical cardiac monitoring}} &
\multicolumn{1}{c|}{\textbf{Ubiquitous sensors}} &
\multicolumn{1}{c|}{\textbf{Setup-free across environments}} &
\multicolumn{1}{c|}{\textbf{Generalizes across hearables}} \\
\hline
Ha et al. (2020)~\cite{ha2020contactless} & mmWave radar & \multicolumn{1}{c|}{\cmark} & \multicolumn{1}{c|}{\xmark} & \multicolumn{1}{c|}{\xmark} & \multicolumn{1}{c|}{N/A} \\
\hline
Fan et al. (2021)~\cite{fan2021headfi} & Hearable mic & \multicolumn{1}{c|}{\xmark} & \multicolumn{1}{c|}{\cmark} & \multicolumn{1}{c|}{\cmark} & \multicolumn{1}{c|}{\cmark} \\
\hline
Gilliam et al. (2022)~\cite{gilliam2022ear} & Hearable infrasonic mic & \multicolumn{1}{c|}{\xmark} & \multicolumn{1}{c|}{\xmark} & \multicolumn{1}{c|}{\cmark} & \multicolumn{1}{c|}{\xmark} \\
\hline
Fan et al. (2023)~\cite{fan2023apg} & Hearable mic and speaker & \multicolumn{1}{c|}{\xmark} & \multicolumn{1}{c|}{\cmark} & \multicolumn{1}{c|}{\cmark} & \multicolumn{1}{c|}{\xmark} \\
\hline
Butkow et al. (2023)~\cite{butkow2023heart} & Hearable in-ear mic & \multicolumn{1}{c|}{\xmark} & \multicolumn{1}{c|}{\cmark} & \multicolumn{1}{c|}{\cmark} & \multicolumn{1}{c|}{\xmark} \\
\hline
Cao et al. (2024)~\cite{cao2024earsteth} & Hearable mic & \multicolumn{1}{c|}{\xmark} & \multicolumn{1}{c|}{\cmark} & \multicolumn{1}{c|}{\cmark} & \multicolumn{1}{c|}{\xmark} \\
\hline
Chen et al. (2024)~\cite{chen2024exploring} & Over-ear headphone speaker & \multicolumn{1}{c|}{\xmark} & \multicolumn{1}{c|}{\cmark} & \multicolumn{1}{c|}{\cmark} & \multicolumn{1}{c|}{\cmark} \\
\hline
Li et al. (2024)~\cite{li2024ecg} & Surface acoustic wave microphone & \multicolumn{1}{c|}{\xmark} & \multicolumn{1}{c|}{\xmark} & \multicolumn{1}{c|}{\cmark} & \multicolumn{1}{c|}{N/A} \\
\hline
Fu et al. (2025)~\cite{fu2025enabling} & Hearable IMU & \multicolumn{1}{c|}{\cmark} & \multicolumn{1}{c|}{\xmark} & \multicolumn{1}{c|}{\cmark} & \multicolumn{1}{c|}{\xmark} \\
\hline
\red{Islam et al. (2025)}~\cite{islam2025ballistobud} & \red{Hearable IMU and PPG sensor} & \multicolumn{1}{c|}{\xmark} & \multicolumn{1}{c|}{\xmark} & \multicolumn{1}{c|}{\cmark} & \multicolumn{1}{c|}{\xmark} \\
\hline
Liu et al. (2025)~\cite{liu2025earmeter} & Hearable mic & \multicolumn{1}{c|}{\xmark} & \multicolumn{1}{c|}{\cmark} & \multicolumn{1}{c|}{\cmark} & \multicolumn{1}{c|}{\xmark} \\
\hline
Christofferson et al. (2025)~\cite{christofferson2025artearial} & Hearable mic & \multicolumn{1}{c|}{\xmark} & \multicolumn{1}{c|}{\cmark} & \multicolumn{1}{c|}{\cmark} & \multicolumn{1}{c|}{\xmark} \\
\hline
\textbf{{\sysname} (ours)} & \textbf{Hearable mics and speakers} &
\multicolumn{1}{c|}{\textbf{\cmark}} &
\multicolumn{1}{c|}{\textbf{\cmark}} &
\multicolumn{1}{c|}{\textbf{\cmark}} &
\multicolumn{1}{c|}{\textbf{\cmark}} \\
\hline
\end{tabular}%
}

\caption{{\bf Comparison of representative cardiac sensing systems.} Representative works from the past five years that focus on biosignal reconstruction for mobile and wireless devices. {\sysname} is designed to extract micro-mechanical cardiac vibrations (SCG and GCG) across a range of hearables.}
\vspace{-2em}
\label{tab:comparison}
\end{table*}

\noindent {\bf Cardiac signals at the ear.} When the heart beats, it produces sounds in particular the characteristic ``lub-dub'' sound of heart valves closing which can be measured at the chest using a stethoscope in a process known as phonocardiography (PCG). Beyond these audible sounds, the heart's mechanical activity generates minute chest wall motions that can be measured through seismocardiography (SCG)~\cite{zanetti2013seismocardiography} and gyrocardiography (GCG)~\cite{jafari2017gyrocardiography}. 

SCG captures fine-grained linear vibrations of the chest wall, while GCG provides complementary information on rotational dynamics. Unlike PCG, which is limited to audible valve closures, SCG and GCG sense both audible and sub-audible mechanical events of the heart. These include clinically important but acoustically silent events such as aortic valve opening, isovolumetric moment, or rapid ejection. Together, SCG and GCG extend the capabilities of health assessments beyond PCG and can be used in the assessment of arrhythmias, myocardial infarction, ischemia, and hemorrhage~\cite{iftikhar2018multiclass,pandia2012extracting,pankaala2016detection,sahoo2018design,salerno1991seismocardiography,solar2017classification,tadi2016real,taebi2017grouping,tavakolian2010characterization,tavakolian2010estimating,zakeri2016analyzing}, and in some cases have even outperformed the electrical signals of the heart (ECG) in detecting coronary heart disease during exercise~\cite{wilson1993diagnostic}. 

Prior work~\cite{cao2024earsteth} has shown that ear-based cardiac sounds can be used to reconstruct the PCG signal. As the PCG signals align closely with the SCG and GCG signal in magnitude ratios and temporal trends (Fig.~\ref{fig:physiology}), this supports the view~\cite{voyatzoglou2022introduction} that the two are coupled views of the same underlying cardiac mechanics. This also suggests that cardiac sounds recorded at the ear can be used to reconstruct the SCG and GCG signal.

\noindent {\bf Prior work on cardiac monitoring with mobile devices.} 
While prior work on hearables have been designed for cardiac monitoring ~\cite{cao2024earsteth,butkow2023heart,fan2023apg,li2024ecg,chen2024exploring,fan2021headfi,gilliam2022ear,liu2025earmeter,christofferson2025artearial} they are mostly focused on reconstructing PCG, ECG and audioplethysmography (APG, which is similar to PPG) signals, arterial pressure waveforms, and respiration volume from in-ear cardiac sounds, we present the first system to reconstruct micro-mechanical cardiac signals across a range of hearables, from ear-based cardiac sounds and extract the timing of micro-cardiac events.

\red{Recent IMU-based approaches including BallistoBud~\cite{islam2025ballistobud} and TWSCardio~\cite{fu2025enabling} have demonstrated the feasibility of capturing cardiac motion using earbud accelerometers. A limitation of this approach is that IMUs are specialized sensors found mainly in true wireless stereo earbuds~\cite{fu2025enabling} or high-end hearables~\cite{chen2024exploring}. In contrast, our approach reconstructs fine-grained cardiac waveforms from widely available in-ear microphones and speakers, which is an approach that applies more broadly across even low-end hearables.}


Smartphone IMU–based approaches~\cite{wang2018seismo,lee2021discrimination,hossein2024smartphone} have measured SCG by strapping the phone to the chest or holding it in place. However, this approach is uncomfortable and this method is not practical for long-term monitoring across multiple hours.

Radar-based systems~\cite{ha2020contactless} have leveraged mmWave to capture chest vibrations corresponding to SCG. This approach requires users to remain directly in front of the radar within an operational range of 25--50~cm, and relies on comparatively expensive devices ($\approx\$350$~\cite{ti_radar}). Because the radar is typically mounted in a fixed position, it requires users to carry and reposition it wherever they want to take a measurement. In contrast, our approach is inherently mobile and can be worn throughout daily life across different environments, and with greater flexibility on user position.

\noindent {\bf Relation to active sonar approaches.} Prior work has explored active sonar techniques using inaudible ultrasonic signals. APG~\cite{fan2023apg} captures a PPG-like signal from earbuds by transmitting tones in the 30--39~kHz range. However, a limitation of this approach is that most commercial earbuds cannot transmit above 30~kHz. A related system, EarMonitor~\cite{sun2023earmonitor}, uses a 16--21~kHz chirp for heart rate monitoring, but the lower end of this range remains audible to young adults~\cite{jilek2014reference}, potentially causing irritation.

We note that an advantage of ultrasonic active sonar approaches is that they can operate during music playback, as their probing signals lie outside the frequency range of typical music. However, our passive sensing system is also able to operate during music playback as SCG and GCG signals occupy low frequencies (< 50~Hz) that lie outside the dominant energy bands of music and environmental sound.


\section{System design}
\label{sec:variability}

\subsection{\red{Designing for accessible, everyday use}}

\red{Our design was guided by principles of accessibility, comort, and everyday usability. We prioritized a hardware-agnostic approach that works across a wide range of hearables including low-end ones to reduce adoption barriers and cost. Leveraging the built-in speaker and microphone enables passive monitoring without requiring users to wear dedicated sensors or require active user participation, which lower the barrier to adoption particularly for older adults who may have cognitive impairment or have lowered technology literacy rates. The ear location was chosen for its consistent coupling and comfort during long-term wear. These considerations align with prior HCI work emphasizing unobtrusive, low-effort health monitoring and sustainable engagement~\cite{karkar2017tummytrials,mariakakis2018drunk}.}






\subsection{Conventional approach of measuring micro-cardiac signals}
\begin{figure}
\centering
\includegraphics[keepaspectratio, width=\linewidth]
    {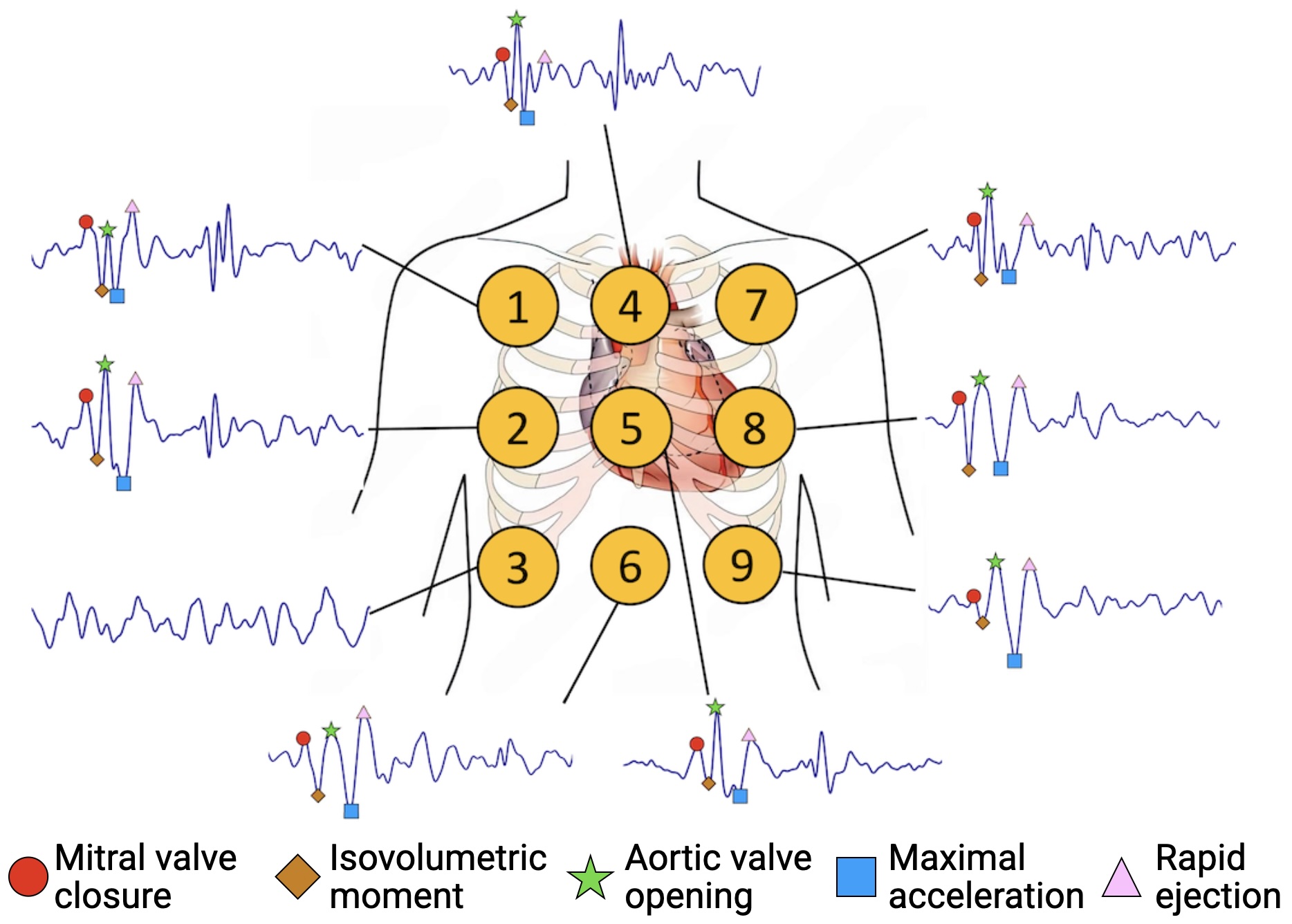}
    \vspace{-1em}
  \caption{{\bf Challenge of conventional IMU-based micro-cardiac measurements.} Differences in sensor placement lead to variations in waveforms, making comparisons across repeated measurements challenging. Precise and consistent placement is difficult to ensure when measurements are performed by lay users outside clinical settings. Each waveform corresponds to a cycle of 800 ms, and amplitudes are normalized to their own maximum.}
  \label{fig:positions}
    \vspace{-1.5em}
    \Description{This figure shows the conventional Inertial Measurement Unit (IMU)-based micro-cardiac measurement results from nine different positions on the human chest, arranged in a grid of three rows and three columns. An outline of a human torso is shown with nine yellow circles, numbered from 1 to 9, placed on the chest area. Position 1 is top-left, 2 is middle-left, 3 is bottom-left, and so on, filling the grid. Each position has a corresponding waveform displayed nearby, representing the measured cardiac signal. These waveforms are complex and show variations in shape and amplitude depending on the measurement location. Five different fiducial points are marked on each waveform, each represented by a unique shape and color, including: a red circle for Mitral valve closing, an orange diamond for Isovolumetric contraction, a green star for Aortic opening, a blue square for Maximal acceleration, and a pink triangle for Rapid Ejection. The diagram highlights how the signal morphology and the presence of these fiducial points vary significantly across different chest locations, with some points being less discernible in certain positions, such as the bottom-left. This illustrates the challenge of obtaining consistent and clear cardiac signals with conventional IMU sensors.}
\end{figure}

Conventionally, the SCG and GCG signal is measured using standalone or smartphone IMUs~\cite{hossein2024smartphone,zanetti2013seismocardiography}. In a clinical setting, the optimal placement has been shown to be between the third and fifth intercostal space~\cite{voyatzoglou2022introduction,sandler2020documenting}. However, it is challenging to ensure that lay users can perform these precise readings repeatedly.

To underscore the importance of placement accuracy, we investigate the impact of smartphone position on SCG waveform variation.
To do this, we placed the smartphone at nine different locations on the torso as shown in Fig.~\ref{fig:positions}. Placements 1 and 2 were on the right lateral ribcage, placements 7 and 8 on the left lateral ribcage, placements 4 and 5 on the sternum, and placements 3, 6, and 9 on the lower abdomen. More specifically, placements 2, 5, and 8 were located between the 3rd and 5th intercostal spaces near the sternum.

Our results show that the SCG signal varies markedly across these nine positions, with an average Pearson correlation of $0.37 \pm 0.34$ between mean cardiac cycles. While most locations (8 out of 9) produced visually recognizable SCG signals, their morphologies differed. At placement 3, no clear cardiac events could be identified, likely because it was the site farthest from the heart.

We then considered five key fiducial points reflecting micro-cardiac events from medical literature~\cite{di2013wearable,gurev2012mechanisms}: mitral valve closure (MC), isovolumetric moment (IM), aortic valve opening (AO), maximal acceleration (MA), and rapid ejection (RE). We observed that at sternum locations (4 and 5) and upper torso sites (1, 4, 7), events such as AO and IM appeared sharp, crisp, and stereotyped, exhibiting well-defined peaks. In contrast, at the abdomen or lower ribs, waveforms became broader and more attenuated; some cardiac events appeared smeared, with delayed or ambiguous onsets. These variations are due to differences in channel propagation where tissue, muscle, and fat create time delays and frequency-dependent attenuation. These results are supported by prior work~\cite{sandler2020documenting} which have shown that IMU-based SCG measurements are highly sensitive to placement.

This analysis reveals a limitation of current approaches for micro-mechanical cardiac measurements outside the clinic, and points to the potential advantage of hearables, which maintain a relatively fixed position in the ear and are therefore less sensitive to placement variation. We examine this benefit in further detail in the next subsection.

\subsection{Characterizing cardiac signals measured on mobile devices.}
The goal of this initial set of experiments is to assess the robustness of ear-based cardiac sounds, SCG, and GCG signals to three different real-world conditions: (1) repeated remounting of devices, (2) inter-user differences, and (3) inter-device differences. We quantify the variability introduced by these factors and describe calibration strategies to mitigate their impact.


\subsubsection{Hearables for capturing cardiac sounds}

\begin{figure}
\centering
\includegraphics[keepaspectratio, width=\linewidth]
    {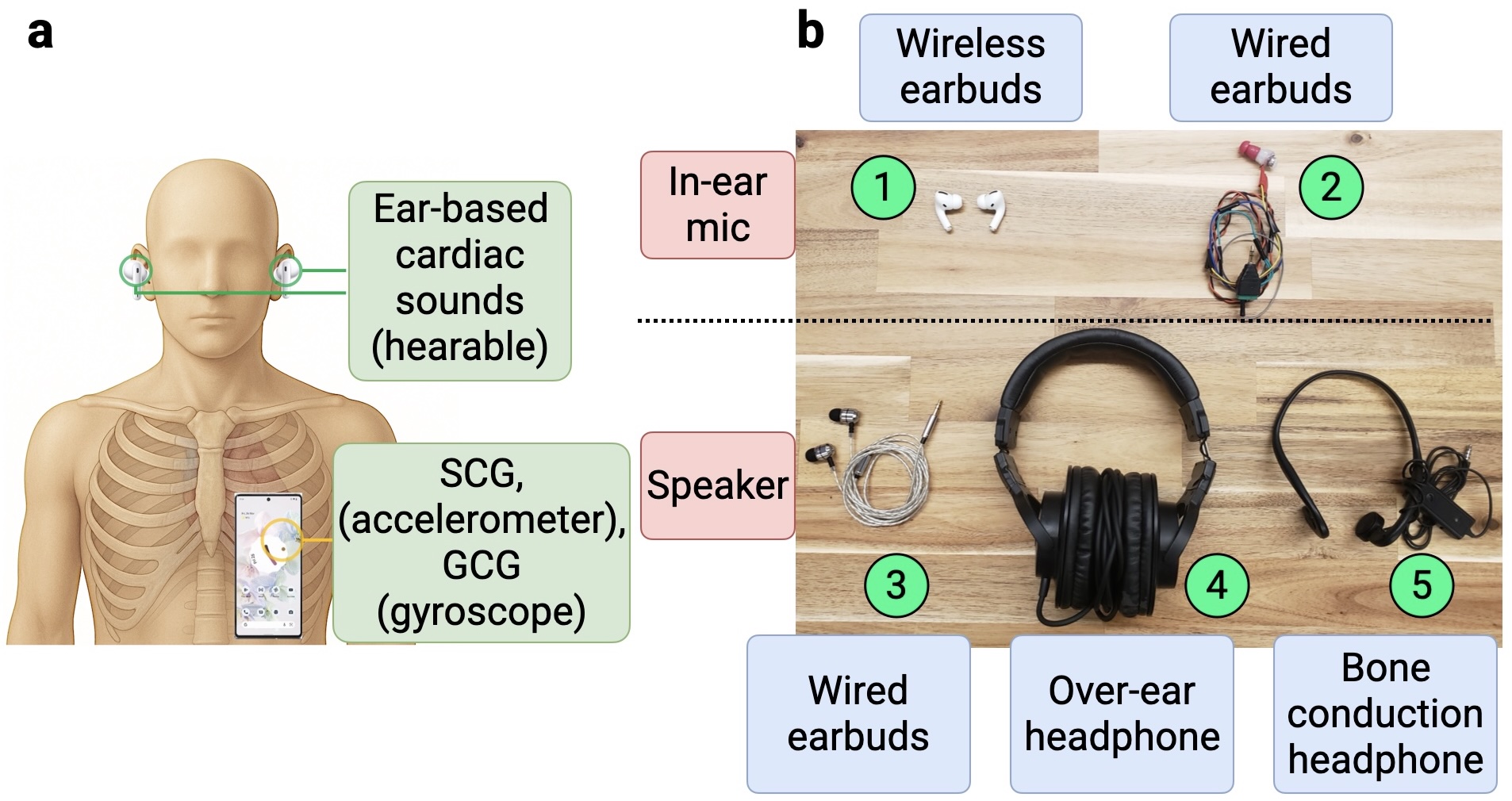}
    \vspace{-1em}
  \caption{{\bf Dataset collection setup. (a)} Ear-based cardiac sounds are measured using the microphone or speaker of a hearable; mechanical cardiac vibrations are measured at the left lower sternal border around the heart using a smartphone IMU. {\bf (b)} Hearables used for data collection span a range of device types.}
  \label{fig:wearing_devices}
  \Description{The two sub-figure illustrate the dataset collection setup of the LubDubDecoder system. The left sub-figure(a) shows a diagram of a human torso with a hearable device placed in their ear. The diagram indicates that the earable collects ear-based cardiac sounds. A second component, a smartphone, is positioned on the person's chest. This smartphone, equipped with an accelerometer and gyroscope, is used to simultaneously capture SCG and GCG signals, serving as ground truth data. The right sub-figure(b) displays the five different types of hearable devices used in the study, organized into two rows. The top row, labeled "In-ear mic," contains two examples: (1) a pair of white wireless earbuds and (2) a pair of wired earbuds. The bottom row, labeled "Speaker," features three devices: (3) a pair of wired earbuds, (4) a large over-ear headphone, and (5) a bone-conduction headphone.}
\end{figure}

\begin{figure*}
\centering
\includegraphics[keepaspectratio, width=\linewidth]
    {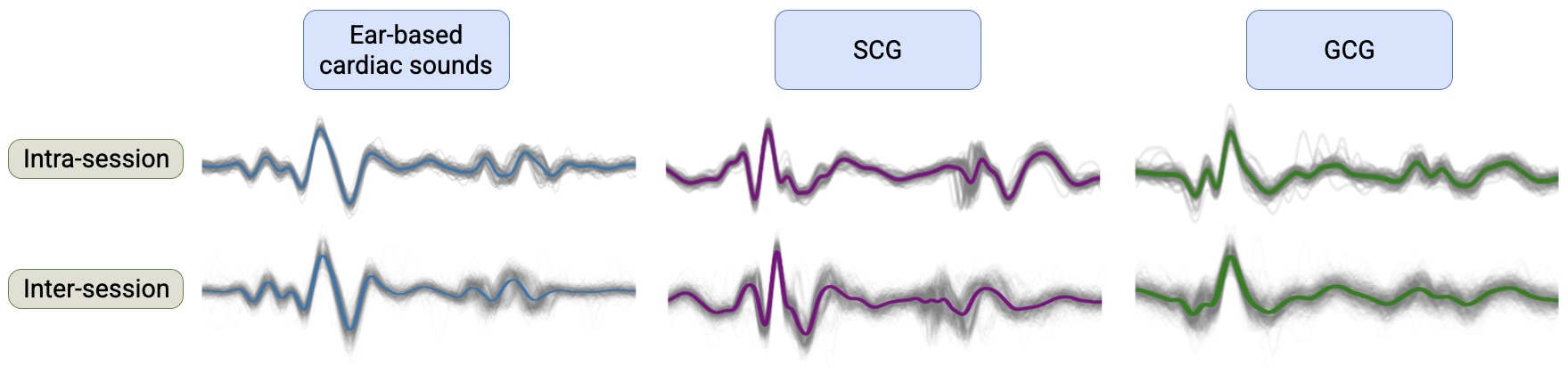}
  \caption{{\bf Effect of device remounting on cardiac signals.} Within a single session, cardiac signals show similar morphology across cycles ($n=60$ cycles). After remounting the hearable and smartphone, ear-based cardiac sounds and micro-mechanical signals maintain comparable waveform shapes, with a modest increase in variability across cycles. Colored opaque line is mean signal across cardiac cycles, all cardiac cycles are overlaid in translucent gray. Each waveform corresponds to a cycle of 800~ms, and amplitudes are normalized to their own maximum.}
  \label{fig:remounting_avg_overlay}
    \Description{This figure illustrates the effects of device remounting on signal consistency for three types of cardiac signals: Ear-based cardiac sounds, Seismocardiogram (SCG), and Gyrocardiogram (GCG). The figure is arranged in a grid with two rows and three columns. The top row is labeled "Intra-session," showing signal consistency within a single recording session, while the bottom row is labeled "Inter-session," demonstrating consistency after the device has been removed and re-worn. Each of the three columns represents a different signal type. In the first column, ear-based cardiac sound waveforms are shown in blue, with a single darker blue line representing the mean signal and all other individual cycles overlaid in a translucent gray. The second column shows SCG signals in purple, and the third column displays GCG signals in green, following the same overlay pattern. Each waveform corresponds to a cycle of 800 ms, and amplitudes are normalized to their own maximum.}
\end{figure*}
\begin{table*}[t]
\centering
\setlength{\tabcolsep}{4pt}
\renewcommand{\arraystretch}{1.05}

\begin{minipage}[t]{0.48\textwidth}
\centering

\begin{tabular}{@{}p{2.2cm}p{2.1cm}p{1.6cm}p{1.6cm}@{}}
\toprule
 & \multicolumn{1}{c}{\textbf{Ear sounds}} & \multicolumn{1}{c}{\textbf{SCG}} & \multicolumn{1}{c}{\textbf{GCG}} \\
\midrule
\textbf{Intra-session} & \multicolumn{1}{c}{0.88 $\pm$ 0.08} & \multicolumn{1}{c}{0.89 $\pm$ 0.06} & \multicolumn{1}{c}{0.87 $\pm$ 0.06} \\
\midrule
\textbf{Inter-session} & \multicolumn{1}{c}{0.82 $\pm$ 0.09} & \multicolumn{1}{c}{0.73 $\pm$ 0.19} & \multicolumn{1}{c}{0.79 $\pm$ 0.07} \\
\midrule
\textbf{Inter-user} & \multicolumn{1}{c}{0.31 $\pm$ 0.19} & \multicolumn{1}{c}{0.03 $\pm$ 0.24} & \multicolumn{1}{c}{-0.03 $\pm$ 0.36} \\
\midrule
\textbf{Inter-device} & \multicolumn{1}{c}{0.15 $\pm$ 0.36} & \multicolumn{1}{c}{0.22 $\pm$ 0.19} & \multicolumn{1}{c}{0.42 $\pm$ 0.29} \\
\bottomrule
\end{tabular}
\captionof{table}{\textbf{Cardiac waveform variability across different experimental conditions.} Waveform variability is measured using Pearson correlation coefficient (mean $\pm$ std). Ear sounds are recorded from the in-ear speaker. SCG and GCG are recorded from the smartphone IMU against the chest.}
\label{tab:variability1}
\end{minipage}
\hfill
\begin{minipage}[t]{0.50\textwidth}
\centering
\begin{tabular}{@{}l l
    >{\centering\arraybackslash}p{1.4cm}
    >{\centering\arraybackslash}p{1.4cm}
    >{\centering\arraybackslash}p{1.4cm}
    >{\centering\arraybackslash}p{1.4cm}@{}}
\toprule
 &  &
 \parbox[c]{1.4cm}{\centering\textbf{Intra\\session}} &
 \parbox[c]{1.4cm}{\centering\textbf{Inter\\session}} &
 \parbox[c]{1.4cm}{\centering\textbf{Inter\\user}} &
 \parbox[c]{1.4cm}{\centering\textbf{Inter\\device}} \\
\midrule
\multirow{4}{*}{\textbf{SCG}}
 & \textbf{AO--MC} & 2.6 & 6.7 & 29.3 & 31.2 \\
 & \textbf{AO--IM} & 1.4 & 2.6 & 18.8 & 19.7 \\
 & \textbf{AO--MA} & 5.0 & 3.5 & 13.1 & 14.6 \\
 & \textbf{AO--RE} & 7.6 & 9.1 & 26.2 & 27.6 \\
\midrule
\multirow{4}{*}{\textbf{GCG}}
 & \textbf{AO--MC} & 2.2 & 8.5 & 27.8 & 31.7 \\
 & \textbf{AO--IM} & 3.9 & 5.5 & 13.8 & 18.4 \\
 & \textbf{AO--MA} & 6.8 & 1.7 & 24.8 & 12.4 \\
 & \textbf{AO--RE} & 7.3 & 7.2 & 29.7 & 23.9 \\
\bottomrule
\end{tabular}
\captionof{table}{\textbf{Timing variability of fiducial point timing across different experimental conditions.} Standard deviations (ms) for timing from aortic opening (AO) to four fiducial points in SCG and GCG from the smartphone IMU against the chest.}
\label{tab:variability2}
\end{minipage}
\end{table*}

Our system measures ear-based cardiac sounds through different hearables which can be divided into three different form factors {\bf (1) in-ear earbuds, (2) over-ear headphones,  and (3) bone conduction headphones}, that each receive the heart sounds in different ways.

\squishlist
\item {\bf In-ear earbuds.} The in-ear microphones and speakers on earphones are able to detect cardiac sounds due to the occlusion effect~\cite{stone2014technique}, a phenomenon that amplifies low-frequency body sounds when the ear canal is sealed. The rubber ear tips of commercial earbuds naturally create this seal and have been shown to capture sounds from the heart.

\item {\bf Over-ear headphones.} For the over-ear headphones which typically only contain a speaker, we leverage the principle of acoustic reciprocity which shows that the speaker diaphragm can work in reverse and be used to act as a microphone to measure the cardiac sounds~\cite{chen2024exploring}. It additionally takes advantage of the passive noise isolation of the ear cup design to pick up the subtle signals.

\item {\bf Bone conduction earphones.} The bone conduction earphones are positioned at the tragus of the ear and measure the lower-frequency cardiac sounds from pulse waves which causes tissue compression and subtle deformations of the skin. We note that these sounds have been picked up by surface acoustic wave microphones at the carotid artery (neck area)~\cite{ecg}.
\squishend

To learn the mapping between the ear-based cardiac sounds, and the micro-mechanical cardiac events at the heart, our system collects paired cardiac sounds at the ear through a hearable, and the micro-mechanical signal through a smartphone IMU. Details of the hearables and smartphone used and their data acquisition settings are described in {\bf Sec.~\ref{sec:human_subjects}}.

\subsubsection{Measuring waveform variation}

We characterize waveform variation using two metrics:

\squishenum
\item \textbf{Waveform similarity.} We use the Pearson correlation coefficient to quantify waveform similarity, computing it across pairwise comparisons of cardiac cycles within a stream, across sessions, and across users or devices. The coefficient ranges from –1 to 1, where –1 indicates a perfect negative correlation and 1 indicates a perfect positive correlation.

\item \textbf{Stability of fiducial timing in SCG and GCG signals.}
   For SCG and GCG, we use the AO as the anchor point as it is often the most distinct micro-cardiac event, and we measured the temporal distances to the other four events and assessed their variability across cycles. 
\squishenumend

Our algorithms for signal segmentation, waveform reconstruction, and fiducial point labeling are described in greater detail in {\bf Sec.~\ref{sec:pipeline}}.

\begin{figure*}
\centering
{\includegraphics[keepaspectratio, width=\linewidth]
    {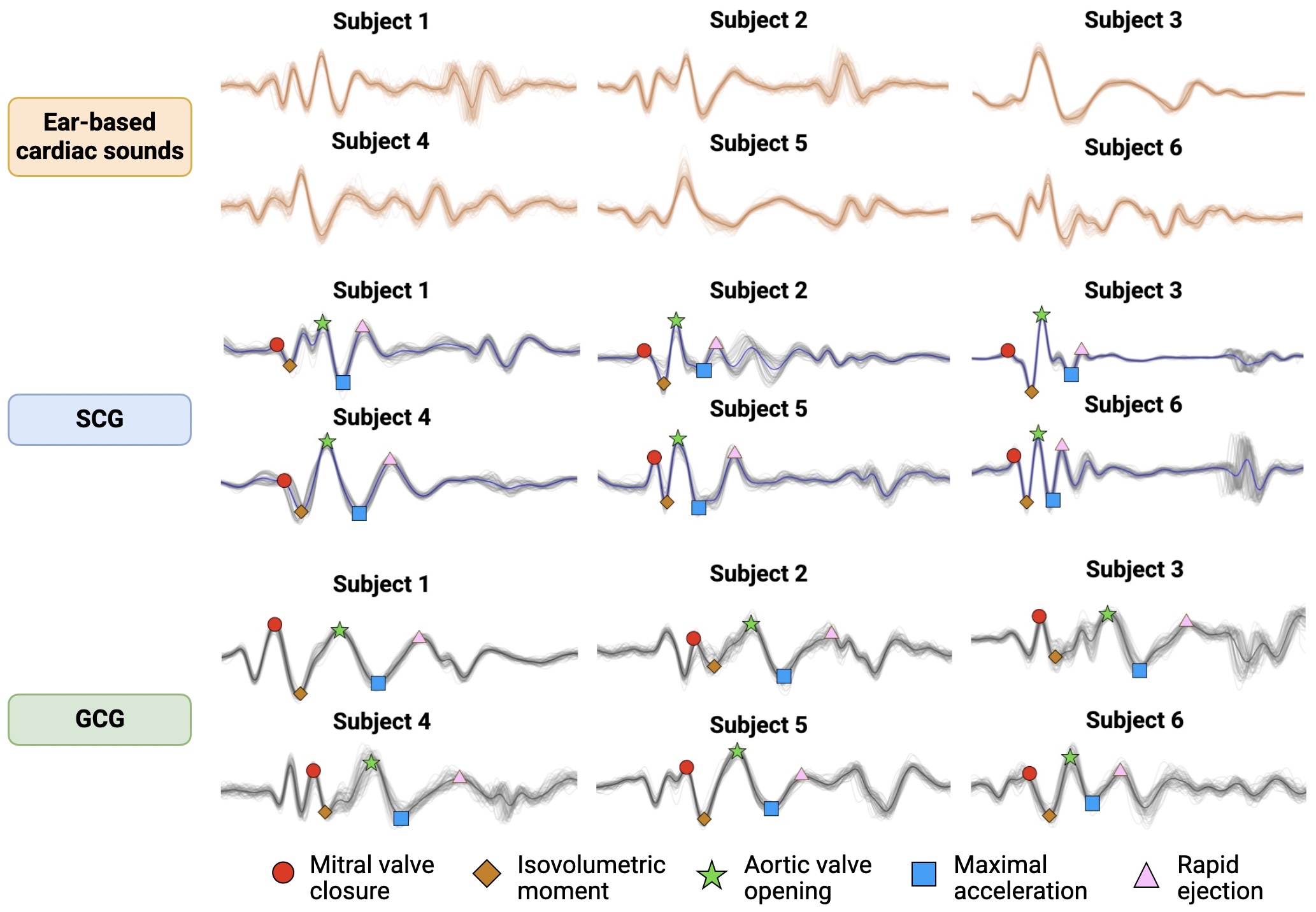}}
  \caption{{\bf Effect of individual physiology on cardiac signal variability.} Waveform variability is presented for a random subset of $n=6$ subjects from our human subjects study showing differences in ear-based cardiac sounds, SCG, and GCG signals ($n=60$ cycles). Solid opaque lines represents the mean across all cardiac cycles within each subject, all cardiac cycles are overlaid in translucent color. Each waveform corresponds to a cycle of 800 ms, and amplitudes are normalized to their own maximum.}
  \label{fig:scg_diversity}
  \vspace{-1em}
  \Description{This figure illustrates the waveform diversity of three different cardiac signals—ear-based cardiac sounds, Seismocardiogram (SCG), and Gyrocardiogram (GCG)—across six different subjects. The figure is organized into three main rows, each corresponding to a signal type. The top row, colored in orange, displays the ear-based cardiac sound signals for all six subjects. The middle row, in blue, shows the SCG signals. The bottom row, in a light green, presents the GCG signals. Within each row, there are six sub-panels, one for each subject, numbered from 1 to 6. Each sub-panel contains a central, bolded line representing the mean signal waveform, with numerous individual cardiac cycles overlaid in translucent gray. On the SCG and GCG signals, five key fiducial points are marked to indicate specific events in the cardiac cycle: a red circle for Mitral valve closing, an orange diamond for Isovolumetric contraction, a green star for Aortic opening, a blue square for Maximal acceleration, and a pink triangle for Rapid Ejection. This figure effectively highlights the significant morphological differences in cardiac signals not only between different types of signals but also from person to person.}
\end{figure*}

\begin{figure*}
\centering
\includegraphics[keepaspectratio, width=\linewidth]
    {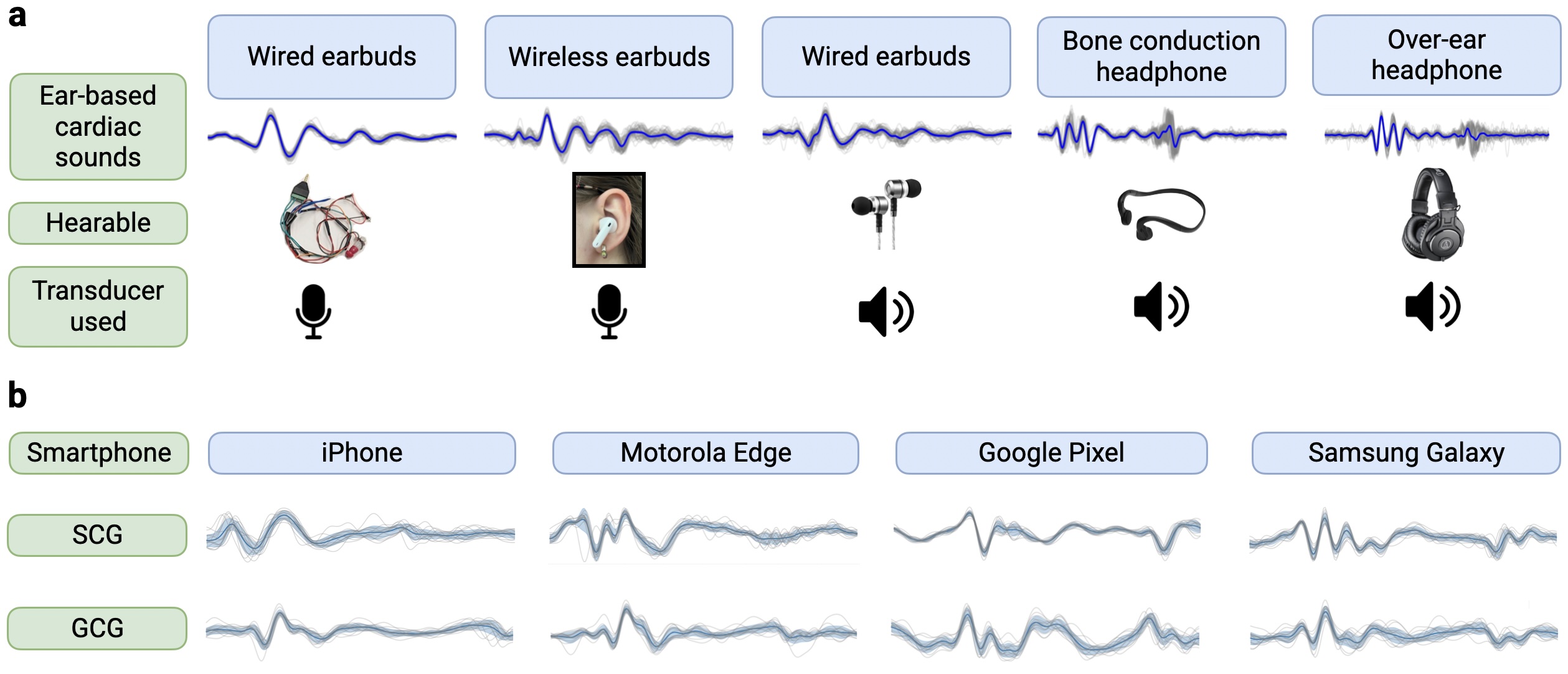}
    \vspace{-1em}
  \caption{{\bf Effect of hardware differences on cardiac signals.} Hardware differences across {\bf (a)} hearables create differences in the measured ear-based cardiac sounds ($n=60$ cycles) and {\bf (b)} across smartphones create differences in the measured SCG and GCG signals ($n=15$ cycles). These differences motivate the need for our zero-effort calibration procedure upon use of a new device. Blue line is the mean signal across cardiac cycles, shaded blue region represents one standard deviation from the mean, all cardiac cycles are overlaid in translucent gray. Each waveform corresponds to a cycle of 800~ms, and amplitudes are normalized to their own maximum.}
  \label{fig:cross_device}
    \vspace{-1em}
    \Description{The two sub-figures illustrate how hardware variations affect the collected cardiac signals. Sub-figure (a), at the top, shows the impact of different hearable devices. It is organized into three rows and five columns. The first row displays a cardiac sound waveform for each hearable. The second row shows images of the five devices: a pair of wired earbuds, a pair of wireless earbuds, a second type of wired earbuds, a bone-conduction headphone, and an over-ear headphone. The third row indicates the transducer type used, with the first two columns showing a microphone icon and the last three columns showing a speaker icon. For all waveforms, a purple line represents the mean signal across multiple heartbeats, with individual beats overlaid in translucent gray. Sub-figure (b), at the bottom, demonstrates the effect of different smartphone models on signals. It has two rows and four columns, with each column representing a different smartphone: iPhone, Motorola Edge, Google Pixel, and Samsung Galaxy. The top row shows the Seismocardiogram (SCG) signal for each phone, while the bottom row shows the Gyrocardiogram (GCG) signal. In these waveforms, a blue line represents the mean signal, the translucent gray lines are individual heartbeats, and a shaded blue region indicates one standard deviation from the mean. All waveforms are normalized in amplitude and represent a single 800-millisecond cardiac cycle.}
\end{figure*}

\noindent {\bf Intra-session.} This analysis provides a baseline for cardiac cycle variability over time within an individual. To characterize this, we had a single participant wear the wireless earbuds and place the smartphone on their chest without removal or re-mounting over the course of 2 minutes. We then segmented the ear-based cardiac sounds, SCG, and GCG signals into individual cardiac cycles of fixed duration (800~ms) corresponding to the typical length of a single cardiac cycle~\cite{widmaier2022vander} (further description of algorithm in {\bf Sec.~\ref{sec:pipeline}}). \rr{We note that this participant was a member of the study team that provided informed written consent prior to any data collection.}

We plot these signals visually in Fig.~\ref{fig:remounting_avg_overlay} which shows that these three cardiac signals retain consistent shape. We quantify this in Table~\ref{tab:variability1}, and show that the mean correlation across cycles was 0.87 $\pm$ 0.11 for ear-based cardiac sounds, 0.86 $\pm$ 0.07 for SCG, and 0.87 $\pm$ 0.06 for GCG. For the timing of the micro-cardiac events in the SCG signal, the mean standard deviations between the AO and the other fiducial points ranged from 1.4 to 7.6~ms (Table~\ref{tab:variability2}).

\noindent {\bf Inter-session.} We next consider how much the signals change when the device is removed and re-mounted. This is an important practical consideration as hearables are frequently taken off and worn again in daily life. To evaluate this, we recruited two participants as before and perform seven remounting trials. In each trial, the hearable was removed and re-mounted, as well as the smartphone for SCG and GCG data collection, followed by a 60-second recording session. \rr{We note that these two participants are members of the study team that provided informed written consent prior to any data collection.}

\textit{As shown in Fig.~\ref{fig:remounting_avg_overlay} all three cardiac signals retain a similar shape for a single user both intra-session and across remounted sessions.} This is quantified across both recruited users with the mean Pearson correlation of the ear-based cardiac sounds across all remounts as 0.82 $\pm$ 0.09 which is slightly lower, but still comparable to the intra-session correlation of 0.87 $\pm$ 0.11 (Table~\ref{tab:variability1}). This stability is likely because hearables occupy a fixed anatomical position at the ear or head, with limited room for placement variability. The SCG signal was more sensitive to remounting. When the smartphone was repositioned on the participant's body at approximately the same location, the waveform similarity decreased to 0.73 $\pm$ 0.19 from the intra-session similarity of 0.86 $\pm$ 0.07.

When repeating the fiducial point timing analysis, the variation in temporal distances between AO and the other fiducial points ranged from 2.6 to 9.1~ms, which is on average 1.3~ms higher than the intra-session variability reported earlier (Table~\ref{tab:variability2}). This increased variability likely arises from small but unavoidable differences in device placement, orientation, and coupling to the body. 

\textit{We evaluate the end-to-end effect of remounting variation on system performance in the evaluation in {\bf Sec.~\ref{sec:benchmark}} and show that the system is able to operate effectively across remounting sessions.}

\noindent {\bf Inter-user.} Beyond temporal and remounting variability, another critical factor is how cardiac signals vary across different individuals. Inter-user variability reflects intrinsic physiological differences that affect how hearables and smartphones capture the signals. To assess this, we analyzed recordings collected from 6 randomly selected subjects from our human subjects study using the same hearable device (in-ear mic from the wireless earbuds). \rr{We describe in {\bf Sec.~\ref{sec:population}} details of the population and note that all participants provided informed written consent prior to any data collection.}

As shown in Fig.~\ref{fig:scg_diversity} across all three cardiac signals, there is substantial waveform morphological diversity. Quantitatively, the waveform similarity exhibited low correlations 0.31 for the ear-based cardiac sounds, and essentially no similarity of 0.03 $\pm$ 0.24 and -0.03 $\pm$ 0.36 for the SCG and GCG waveforms respectively (Table~\ref{tab:variability1}). The fiducial points' distance standard deviations for both the SCG and GCG were in the range 13-30~ms, which is larger than that exhibited within a single individual which was 1-8~ms (Table~\ref{tab:variability2}). This suggests that mechanical event timing is highly individualized, reflecting differences in cardiac physiology. 

\textit{Given intrinsic differences across individual physiologies which affect the heart-to-ear channel, these results suggest that calibration is needed to adapt to physiological differences for new individuals.}

\noindent {\bf Inter-device.} Finally, we examined the influence of device hardware. Even when worn by the same user, different hearable devices introduce variability due to differences in microphone and speaker characteristics in particular sensitivity, and frequency response. Similarly, hardware differences in smartphone IMUs result in variability in the measured SCG and GCG waveform.

To evaluate device-related variability, we recorded ear-based cardiac sounds from three participants, each using five different hearable devices (Fig.~\ref{fig:wearing_devices}), and the SCG and GCG signal across four different smartphones (iPhone 12, Motorola Edge, Google Pixel 7, and Samsung Galaxy S9). \rr{We note that these three participants are members of the study team that provided informed written consent prior to any data collection.} We show the variation in the physiological signals for one of the participants in Fig.~\ref{fig:cross_device} which illustrates that the ear-based cardiac sounds exhibited substantial differences across devices. We then show the quantitative variation for this inter-device variation averaged across three users and show the Pearson correlation for the ear-based cardiac sounds, SCG, and GCG signal was 0.15 $\pm$ 0.36, 0.22 $\pm$ 0.19, and 0.42 $\pm$ 0.29 respectively (Table~\ref{tab:variability1}), which is lower than the intra-session and inter-session similarities. The fiducial point timing variability is also higher, ranging from 12 to 31~ms for the SCG and GCG signals, in comparison to the intra-session variability of 1 to 8~ms (Table~\ref{tab:variability2}). 

\textit{These differences in hardware motivate the need for a device-specific normalization procedure when deploying our method on unseen hardware. Without such normalization, the learned transformation between signals may not generalize, since the same underlying cardiac activity may manifest differently across devices.}

\subsection{Reconstructing micro-cardiac signals}

\begin{figure}[h]
\centering
\includegraphics[keepaspectratio, width=\linewidth]
{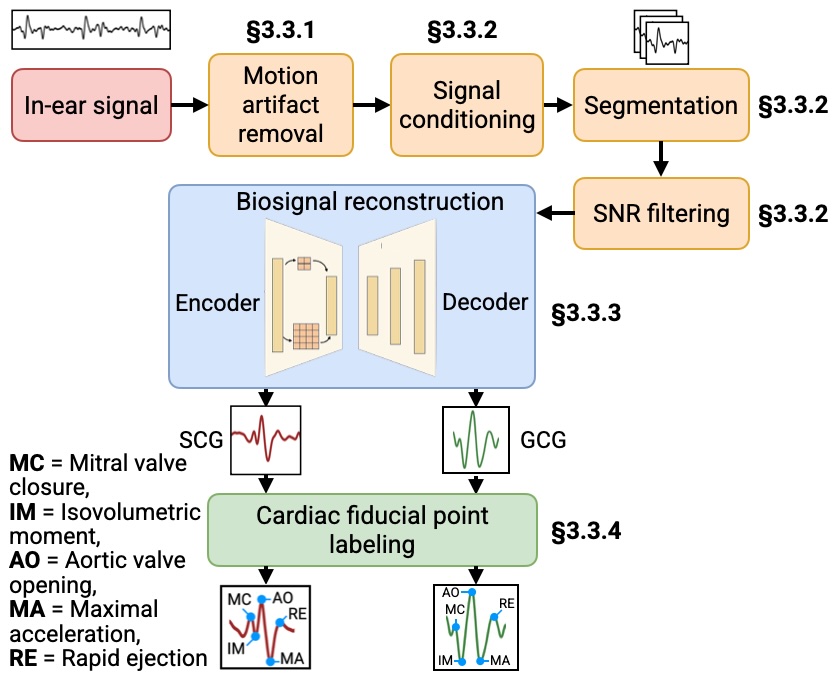}
    \vspace{-1em}
  \caption{{\bf {\sysname} system pipeline.}}
    \vspace{-1em}
  \label{fig:pipeline}
  \Description{This figure is a detailed flowchart illustrating the processing pipeline of the LubDubDecoder system. The process starts at the top left with an In-ear signal, which is represented by a small waveform icon. The signal then flows sequentially through three orange processing blocks: Motion artifact removal(section 3.3.1), Signal conditioning(section 3.3.2), and Segmentation(section 3.3.2), with each step refining the signal. After segmentation, the signals undergo an SNR filtering(section 3.3.2), and those that pass proceed to a large central blue block for Biosignal reconstruction, which is carried out by an Encoder-Decoder neural network(section 3.3.3). The output of this reconstruction is two distinct biosignals: Seismocardiogram (SCG) and Gyrocardiogram (GCG), each represented by its own waveform. Finally, these reconstructed signals are processed in a green block labeled Cardiac fiducial point labeling(section 3.3.4), where key points on the waveforms are identified and marked with blue dots. A legend in the bottom-left of the figure defines these points: MC for Mitral valve closure, IM for Isovolumetric moment, AO for Aortic opening, MA for Maximal blood acceleration, and RE for Rapid ejection.}
\end{figure}

We designed a signal processing and machine learning pipeline to reconstruct the SCG and GCG signals from ear-based cardiac signals (Fig.~\ref{fig:pipeline}).

\subsubsection{Motion artifact removal pipeline}
\label{sec:motion_system}

In practical use, the in-ear cardiac sounds can be corrupted by the subject's body and head movements, which introduce motion artifacts into the audio recording. The challenge with detecting head and body motions such as chewing, nodding, or walking is that they generate low-frequency components, which can only be partially filtered out in the frequency domain. Furthermore, these motion signals are often non-periodic, and their amplitude can be larger than the cardiac signals. 

To address this, we designed a data-driven, temporal segmentation approach to identify and discard segments affected by user movement, rather than relying on band-pass filtering alone. We opted to classify and discard 10-second segments that contain motion which avoids the risk of leaving behind short clean fragments that may be too sparse or unreliable to use. 

We recorded a dataset containing six types of motion events: brow raising, chewing, nodding, talking, and walking. All recordings were collected from a single subject using in-ear microphones. \rr{This subject was a member of the study team who provided informed written consent prior to any data collection.} \red{For each acoustic event, we recorded 10-second segments, a duration that is in line with standard practice in large scale audio datasets such as AudioSet~\cite{gemmeke2017audio}, which uses this duration to capture sufficient acoustic context for a diverse range of events for use in machine learning models. This choice does not limit how frequently our system can output results because (1) a real-time implementation can use a sliding window with a small overlap between windows and (2) our waveform reconstruction model can run in real time and complete inference on a 10 second clip in 19~ms on a smartphone.}

\begin{figure}[h]
  \centering
  \includegraphics[width=\linewidth]{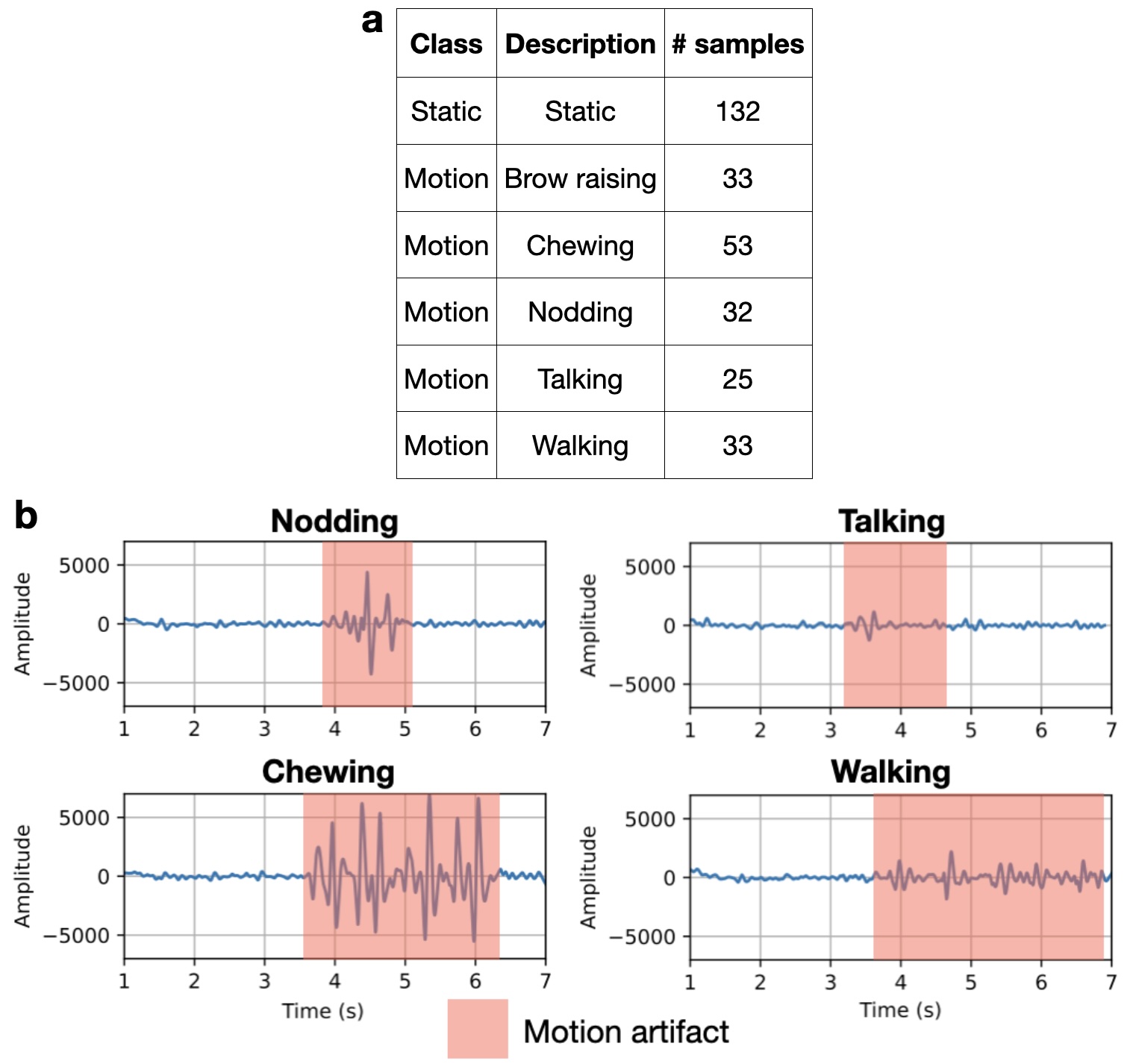}
  \vspace{-2em}
  \caption{{\bf Effect of user motions on in-ear cardiac sounds.} {\bf (a)} Dataset composition used to train our motion artifact removal classifier, showing the number of 10-second audio recordings in each class. {\bf (b)} Motion artifacts produce signals with much higher amplitude than heart sounds, obscuring them. Our system automatically detects and discards segments affected by such artifacts.}
  \label{fig:waveforms}
  \Description{This figure consists of a table and a series of four plots, illustrating effect of user motions on in-ear cardiac sounds. Sub-figure 'a' is a table detailing the composition of the dataset. It has three columns: "Class," "Description," and "# samples." The table shows two main classes: "Static" with a description of "Static" and 132 samples, and "Motion" which is further broken down by "Description" into "Brow raising" (33 samples), "Chewing" (53 samples), "Nodding" (32 samples), "Talking" (25 samples), and "Walking" (33 samples). Sub-figure 'b' is a four-panel grid illustrating the effects of different types of user motion on a signal waveform, with each plot showing a distinct motion artifact. Each of the four plots has an x-axis labeled "Time (s)" from 1 to 7 and a y-axis labeled "Amplitude" from -5000 to 5000. In each plot, a section of the waveform is highlighted by a translucent red box, representing the period of motion. The top-left plot is labeled "Nodding," and the highlighted section shows a period of sharp, high-amplitude, high-frequency signal peaks. The top-right plot is labeled "Talking," and its highlighted section shows a period of relatively lower-amplitude, less distinct signal activity. The bottom-left plot is labeled "Chewing," and the highlighted section shows a long period of very high-amplitude, chaotic signal. Finally, the bottom-right plot is labeled "Walking," and its highlighted section shows a long period of sustained, low-amplitude signal oscillations.}
\end{figure}

We also recorded a set of recordings when the user was static and when there was no external interference or music playback (Fig.~\ref{fig:waveforms}a).  Fig.~\ref{fig:waveforms}b shows that the various motion types produced distinct signal distortions that were clearly visible and would obscure the smaller heart sounds.

To distinguish between segments with and without motion, for each audio segment, we preprocess them using Mel-Frequency Cepstral Coefficients (MFCCs), a widely used feature representation in audio analysis~\cite{gupta2013feature}. MFCCs capture the short-term power spectrum of a signal in a way that reflects how humans perceive sound. In our context, they provided a compact and informative summary of each segment's audio characteristics, which can allow a downstream model to differentiate between silent and motion-contaminated recordings. We evaluate the performance of our motion artifact removal pipeline in the evaluation {\bf Sec.~\ref{sec:benchmark}} and demonstrate it can detect and remove time windows with unwanted motion artifacts.


\subsubsection{Signal processing}\label{sec:pipeline}

We describe the various signal processing steps below to condition the paired ear-based cardiac sounds, SCG and GCG signals for training our biosignal reconstruction model.

\noindent {\bf Signal conditioning.} To ensure that all three cardiac signal streams have consistent dimensions, we first resampled them to 500~Hz. We then applied a 4th-order Butterworth bandpass filter (5--45~Hz) to the signals. This filter choice is consistent with prior literature~\cite{rai2021comprehensive} which notes that the energy of micro-cardiac signals primarily resides below 50~Hz. The lower cutoff at 5~Hz reduces the effects of respiration artifacts, and other low frequency noise. Finally, each segment was z-score normalized (zero mean, unit variance) to minimize the effects of amplitude variability across recordings and improve model training stability.

\noindent {\bf Cardiac cycle segmentation.} To prepare the signals for the model, our goal is to segment the ear-based cardiac sounds and the target SCG and GCG signals into individual cardiac cycles. Segmentation was anchored to the most prominent peak of each modality: the S1 peak for ear-based cardiac sounds and the AO peak for SCG and GCG signals. 

We first identified the S1 and AO anchor points by applying a peak-finding algorithm across the signal while enforcing a minimum spacing of 0.55~s between peaks, which corresponds to a maximum plausible heart rate of 110~bpm, slightly above the upper bound of resting heart rate in healthy adults. For each valid S1 or AO, we then extracted a fixed 800~ms window (200~ms before to 600~ms after the peak), which approximates the duration of a cardiac cycle at a resting heart rate of 75~bpm.

\noindent We additionally apply heuristics specific to each cardiac signal type:

\noindent \textit{(1) Ear-based cardiac sounds.} Each heartbeat produces two heart sounds, S1 and S2. While S1 is generally more prominent, in some users S2 can appear equally strong, leading to ambiguity in peak detection. To differentiate between the two, we compared the average energy in two 400~ms windows placed before and after each detected peak. For true S1 peaks, the right window after the peak would contain the S2 sound and would have higher energy compared to the left window before the peak which would contain background noise. The opposite pattern occurs if S2 peaks were initially detected.

\noindent \textit{(2) SCG and GCG signals.} In SCG and GCG recordings, the AO peak is typically the most prominent feature of each cycle, but inter-subject variability can cause neighboring peaks such as MC or RE to appear equally strong. To improve robustness, we first apply the general peak-finding procedure to identify an initial AO candidate. We then examine all peaks within a narrow 200~ms window centered on this candidate. For each peak, we compute a local prominence score defined as the sum of its amplitude and those of its immediate left and right neighbors. The peak with the highest score is then selected as the definitive AO peak.

\noindent \red{\textbf{Multi-cycle training.} While single-cycle inputs allow the model to learn detailed intra-beat morphology, they provide limited information about the inter-beat features across consecutive heartbeats. To address this, we incorporate a multi-cycle training strategy in which the continuous signal is segmented into 3-second windows that contain multiple cardiac cycles. These longer segments enable the model to capture inter-beat dynamics especially when the user's heart rate changes across time, such as after exercise, or during sleeping. Combined with single-cycle training, this hybrid approach allows the model to learn both fine-grained intra-beat features and broader inter-beat patterns, improving overall reconstruction fidelity.}

\red{To help the model learn inter-beat patterns across a wide range of heart rates, we apply a stretching-based data augmentation strategy~\cite{ha2020contactless}. Specifically, each segment is time-stretched by factors ranging from 0.8 to 1.3, effectively simulating slower and faster heart rates. This augmentation improves the model’s ability to generalize to different heart rates during real-world use.}

\noindent {\bf SNR-guided filtering.} After segmentation, to ensure that only high-quality ear-based cardiac sounds are used for reconstruction, we computed the signal-to-noise ratio for each cardiac cycle and discarded those below a threshold of 7~dB. Because each cardiac cycle contains two distinct heart sounds (S1 and S2), we define a 400~ms window around the S1 peak that captures both S1 and S2 as the signal region, while the remainder of the cycle is treated as background noise. Over a continuous recording, this produces alternating segments of signal and noise. We then compute the power of the signal segments and compare it with that of the noise segments to derive the overall SNR. This measure allows us to evaluate the reliability of each ear’s recording and to select the channel with the higher SNR for subsequent reconstruction tasks.

For each region, the average power was computed as
\[
P = \frac{1}{N}\sum_{i=1}^{N} x[i]^2,
\]
where $x[i]$ are the amplitudes and $N$ is the number of samples in the region.
We denote $P_{\mathrm{signal}}$ and $P_{\mathrm{noise}}$ as the power within the signal and noise regions, respectively.
The SNR is then computed as
\[
\mathrm{SNR} = 10\log_{10}\!\left(\frac{P_{\mathrm{signal}}}{P_{\mathrm{noise}}}\right).
\]

\subsubsection{Model architecture} 
Similar to prior works on biosignal reconstruction~\cite{cao2024earsteth,li2024ecg}, we designed a temporal autoencoder with a multi-branch convolutional architecture to learn a cross-modal mapping from in-ear cardiac signals to SCG waveforms (Fig.~\ref{fig:pipeline}). Our design directly uses raw time-domain signals as input as this preserves the full temporal resolution and allows the model to learn latent structure without compression artifacts associated with transformations like spectrograms and MFCCs.

\noindent {\bf Autoencoder design.} Our model consists of an encoder-decoder structure tailored for cross-modal reconstruction. The encoder transforms the input into a lower-dimensional latent space that captures temporal and physiological features shared between in-ear cardiac sounds and SCG and GCG signals.

\red{The encoder employs a two-branch convolutional neural network with progressively increasing channel dimensionality ($64 \to 128 \to 256 \to 512$). The local branch uses dilated 1D convolutions with small kernel sizes (3 $\times$ 1) to capture fine-grained temporal patterns within individual cardiac cycles, while the global branch employs larger receptive fields (48 $\times$ 1) to model inter-beat context and periodic structure. The outputs from both branches are concatenated and refined through a temporal self-attention module that dynamically reweights feature contributions over time. Each dual-path block is followed by MaxPooling for temporal downsampling and Dropout for regularization. Residual connections and batch normalization layers are integrated throughout to stabilize training and ensure smooth gradient flow.}

\red{Following the encoder, three additional convolutional layers with 512 channels and batch normalization further refine the latent representation, preserving temporal continuity while enhancing local feature selectivity. The decoder mirrors the encoder through four upsampling blocks, each consisting of an upsampling layer (scale factor of 2) followed by a convolutional layer (kernel size 3), batch normalization, and LeakyReLU activation. Channel dimensionality is progressively reduced from $512 \to 256 \to 12 \to 64 \to 32$ to reconstruct the temporal resolution of the original signal. Attention modules are integrated within both the encoder’s dual-path blocks and decoder to enhance temporal coherence and focus on salient cardiac events. Finally, a 1D convolution layer projects the intermediate feature maps into a single output channel representing the reconstructed SCG or GCG waveform.}


We train the model using the Adam optimizer and a mean squared error loss. To improve convergence, a learning rate scheduler is used during training.


\subsubsection{Fiducial point labeling algorithm.} 
We develop an automated pipeline to label fiducial points in SCG and GCG signals. Each cardiac cycle contains five micro-cardiac events that occur in a fixed order: mitral valve closure (MC), isovolumetric moment (IM), aortic valve opening (AO), maximal blood acceleration (MA), and rapid ejection (RE).

After segmenting the waveform into individual cycles, the global maximum within each cycle is identified as the AO peak, which serves as the anchor for locating the remaining fiducial points. The MC point is defined as the maximum peak to the left of AO, and the RE point as the maximum peak to the right. The IM point is selected as the minimum amplitude within the [MC, AO] interval, while the MA point is selected as the minimum amplitude within the [AO, RE] interval.

\subsubsection{Calibration}
\label{sec:calib}

As shown in Table~\ref{tab:variability1} and ~\ref{tab:variability2}, the ear-based cardiac signals, SCG, and GCG waveforms vary across users due to intrinsic physiological differences, and across devices due to hardware variability. This highlights the importance of incorporating calibration strategies to enable more accurate and robust signal reconstruction.

\noindent {\bf Cross-user calibration.} For cross-user generalization, we first train a base model across data from all users in our training cohort. When encountering a new user, we collect a short calibration recording of five cardiac cycles (approximately four seconds) using the smartphone's IMU which is then used to fine-tune the pretrained model, allowing it to adapt to the new user’s physiology with minimal effort. We show in our evaluation in Fig.~\ref{fig:cal_cycle} that five cycles are sufficient to enable strong cross-user performance while keeping calibration effort minimal.


\noindent {\bf Cross-device normalization.}
We leveraged frequency-domain equalization to normalize signals across devices. \textit{For this normalization step, only data from the hearable is needed, not the smartphone. In this way, no explicit user effort is required for this normalization} For each device, we first extracted ten consecutive cardiac cycles (approximately 8 seconds) and computed their mean waveform. Let $x_{\text{ref}}(t)$ and $x_{\text{tgt}}(t)$ denote the mean cycles for the reference and target devices, respectively. Their Fourier transforms are computed using the FFT $\mathcal{F}$:
\begin{equation}
X_{\text{ref}}(f) = \mathcal{F}\{x_{\text{ref}}(t)\}, \quad 
X_{\text{tgt}}(f) = \mathcal{F}\{x_{\text{tgt}}(t)\}.
\end{equation}

We derive a frequency-domain mapping function $H(f)$ by taking the ratio of the two spectra:
\begin{equation}
H(f) = \frac{X_{\text{ref}}(f) \cdot X_{\text{tgt}}^{*}(f)}{|X_{\text{tgt}}(f)|^2 + \epsilon},
\end{equation}
where $(\cdot)^*$ denotes the complex conjugate and $\epsilon$ is a small regularization constant to ensure numerical stability.

Given a new cardiac cycle $x_{\text{tgt}}^{(i)}(t)$ from the target device, we compute its Fourier transform $X_{\text{tgt}}^{(i)}(f)$, apply the mapping $H(f)$, and transform back to the time domain:
\begin{equation}
\hat{x}_{\text{tgt} \rightarrow \text{ref}}^{(i)}(t) 
= \mathcal{F}^{-1} \{ H(f) \cdot X_{\text{tgt}}^{(i)}(f) \}.
\end{equation}

Finally, we apply an energy normalization step to match the amplitude of the reference mean cycle using the L2 norm of the reference and new cardiac cycle:
\begin{equation}
\hat{x}_{\text{norm}}^{(i)}(t) 
= \alpha \cdot \hat{x}_{\text{tgt} \rightarrow \text{ref}}^{(i)}(t), 
\quad \alpha = \frac{\|x_{\text{ref}}(t)\|_2}{\|\hat{x}_{\text{tgt} \rightarrow \text{ref}}^{(i)}(t)\|_2}.
\end{equation}

\section{Feasibility study}
\label{sec:human_subjects}


\subsection{Hardware setup}
We designed a data collection platform (Fig.~\ref{fig:wearing_devices}) that measures ear-based cardiac sounds across five hearables. Two used \textbf{in-ear microphones} and three used \textbf{in-ear speakers}:

\squishenum  
\item Wireless earbuds (Honor Earbuds 3\footnote{https://www.honor.com/global/audio/honor-earbuds-3-pro/}, \textbf{in-ear microphone}, obtained through a collaborative research agreement with Honor Device Co., Ltd.)  
\item Custom wired earbuds (POM-2730L-HD-R\footnote{https://www.digikey.com/en/products/detail/pui-audio-inc/POM-2730L-HD-R/7898330}, \textbf{in-ear microphone})  
\item Wired earbuds (Sephia SP3060\footnote{https://www.amazon.com/sephia-SP3060-Earbuds-Lightweight-Tangle-Free/dp/B0170RBJ9Q}, \textbf{in-ear speaker})  
\item Over-ear headphones (Audio-Technica ATH-M30x\footnote{https://www.amazon.com/dp/B00HVLUQW8}, \textbf{speaker})  
\item Wired bone conduction earphones (FSC Wired Bone Conduction\footnote{https://www.amazon.com/dp/B0915DB6DN}, \textbf{speaker})  
\squishenumend

Audio was sampled at 16~kHz with 16-bit resolution. Audio from the wireless earbuds were streamed in real time via Bluetooth to a host computer. The Babyface AD/DA converter~\cite{babyface} was used to acquire signals from the speaker of the hearables for the purposes of demonstrating proof-of-concept. Subsequent miniaturization efforts could focus on a low-cost integrated PCB design. For convenience, we collected data from a single random ear, however bilateral data collection is possible with the earbuds. A custom smartphone Android app was used to collect the subject's corresponding SCG and GCG data using the onboard IMU. We used the Google Pixel 7 with a sampling rate of 454~Hz. The smartphone was placed at the lower left sternal border between the third and fifth intercostal space where the SCG and GCG signal is the strongest~\cite{voyatzoglou2022introduction,sandler2020documenting}.



\begin{figure*}
\centering
{\includegraphics[keepaspectratio, width=\linewidth]
    {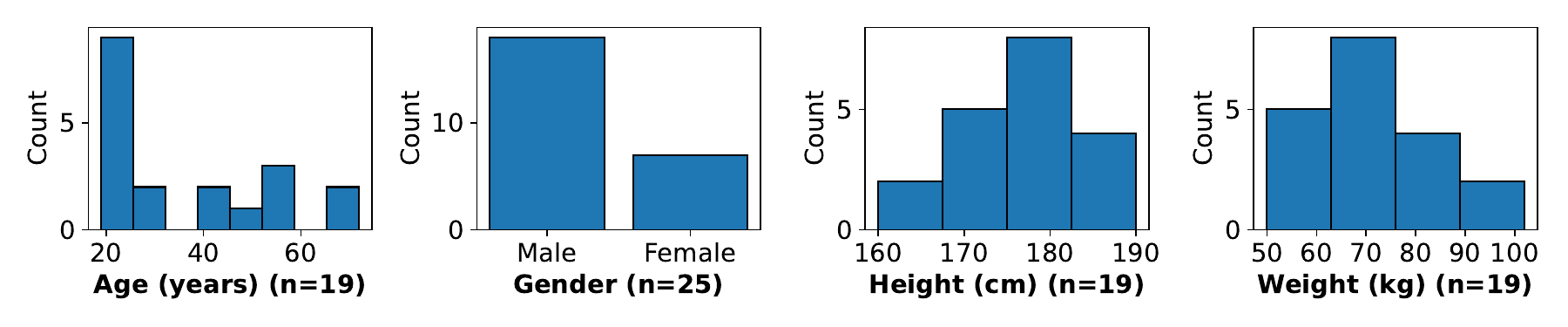}}
  \vspace{-2.5em}
  \caption{{\bf Demographic summary of participants in the human subjects study.} Complete demographic data was collected for \red{19} of \red{25} participants. \red{Gender data was collected for all 25 participants}}
  \vspace{-1em}
  \label{fig:demos}
  \Description{Four histogram panels showing demographic data for our study. Panel 1 (Age) shows the distribution of 19 participants across age ranges from 20 to 70+ years, with the highest count of approximately 7 participants in the 19-25 age range, and smaller counts distributed across older age groups. Panel 2 (Gender) displays a bar chart of 25 participants with 18 males and 7 females. Panel 3 (Height) shows the distribution of 19 participants' heights in centimeters, ranging from 160 to 190 cm. Panel 4 (Weight) shows the distribution of 19 participants' weights in kilograms, ranging from 50 to 100 kg. }
\end{figure*}

\subsection{Dataset collection protocol}
\label{sec:population}
This study was approved by our Institutional Review Board \\ (STUDY2025\allowbreak\_00000155). All studies complied with relevant ethical regulations. Participants were recruited by word of mouth through our university campus. \rr{Informed written consent was obtained for human subjects participating in the study before any study procedures began.} Randomization was not applicable and investigators were not blinded.

Participants above the age of 18 were eligible for the study. Exclusionary criteria include pregnant women, prisoners, adults with cognitive impairments, individuals with allergies or sensitivities to adhesives or sensors, individuals with implanted medical devices, such as pacemakers, individuals with significant mobility impairments. Individuals were screened via inclusion and exclusion criteria prior to the study. We recruited \red{25} adults with a female-to-male ratio of 0.39, of whom \red{19} had complete demographic records: \red{mean age 36.4 ± 17.9 years, height 176.6 ± 7.8~cm, weight 72.0 ± 13.7~kg}. All of whom were healthy and without history of cardiovascular disease (Fig.~\ref{fig:demos}).



We collected data in a quiet room from using the hearables to record ear-based cardiac sounds, and the smartphone IMU for SCG and GCG data. To synchronize the cardiac signals between the hearable and the phone, we tapped both together twice. Participants then remained in a static seated position for 10 minutes as data was recorded. After each measurement session, we aligned the recordings based on the tap events. We collected a total of \red{710} minutes (\red{11.8} hours) of synchronized data from the participants. In a deployment of our system, this synchronization could similarly be achieved by having the smartphone transmit an inaudible acoustic chirp (18--20~kHz) that is simultaneously received by the smartphone and hearable microphones.

For all  \red{25} participants, we collected recordings from the in-ear speaker of wired earbuds, and for 15 of them, also from the in-ear microphone of wireless earbuds. In addition, for a subset of 3 participants recordings were obtained across all five hearables (Fig.~\ref{fig:wearing_devices}).

\section{Evaluation}
\label{sec:eval}
\subsection{Evaluation setup} 
We evaluate the performance of our system through three protocols:\\
\noindent \textit{(1) Within-user evaluation}, where we apply five-fold cross-validation on the waveforms collected for each subject. This essentially captures the upper bound of system performance using a model trained and tested only on data from a single individual.\\
\noindent \textit{(2) Cross-user evaluation}, where a model is pretrained on data from all but one user, fine-tuned on a small number of synchronized cardiac cycles from the hearable and smartphone from the held-out user, and evaluated on the remaining cycles. This calibration step accounts for inter-individual physiological variability that must be adapted for during deployment.\\
\noindent \textit{(3) Cross-device evaluation}, where the model is trained and tested from data measured from the same user, but across five different devices. We first train a model from one reference hearable device. For each unseen hearable device, we use a small number of cardiac cycles to compute normalization weights relative to the reference device’s ear-based heart sounds. These weights are then applied to subsequent cycles from the unseen device, which are passed into the reconstruction model to generate the micro-cardiac signals. We note here, that we only need recordings for the hearable, not the smartphone for the normalization. 

For our within-user and cross-user evaluation, we use signals collected from the in-ear speaker of wired earbuds and in-ear microphone of wireless earbuds (device 3 and 1 in Fig.~\ref{fig:wearing_devices} respectively).

\subsection{Waveform reconstruction performance}
\label{sec:reconstruction}
\begin{figure*}[htbp]
\centering
\begin{subfigure}{\linewidth}
  \centering
  \includegraphics[width=0.99\linewidth]{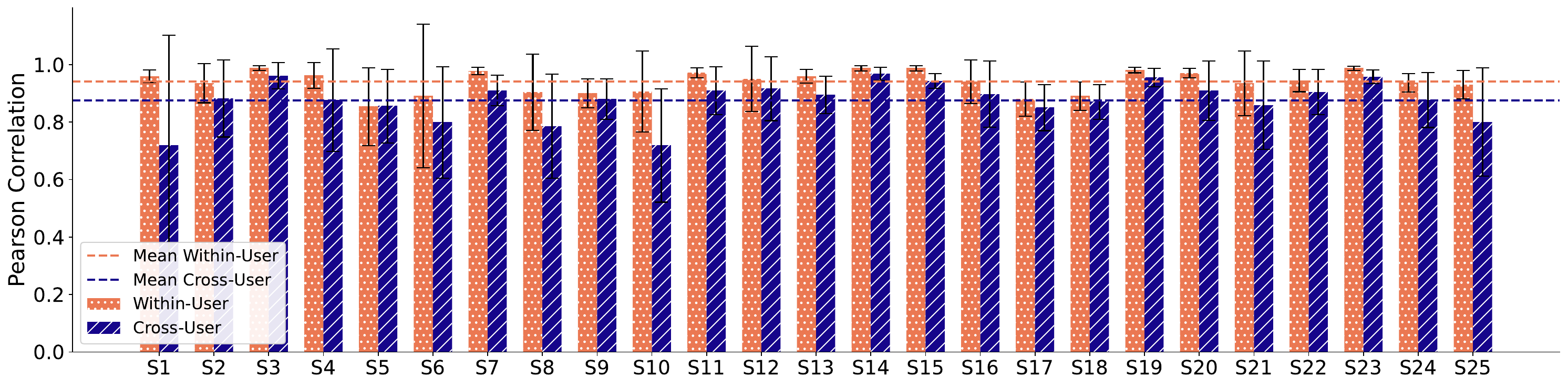}
  \caption{\red{\bf SCG similarity}}
\end{subfigure}

\vspace{1em}

\begin{subfigure}{\linewidth}
  \centering
  \includegraphics[width=0.99\linewidth]{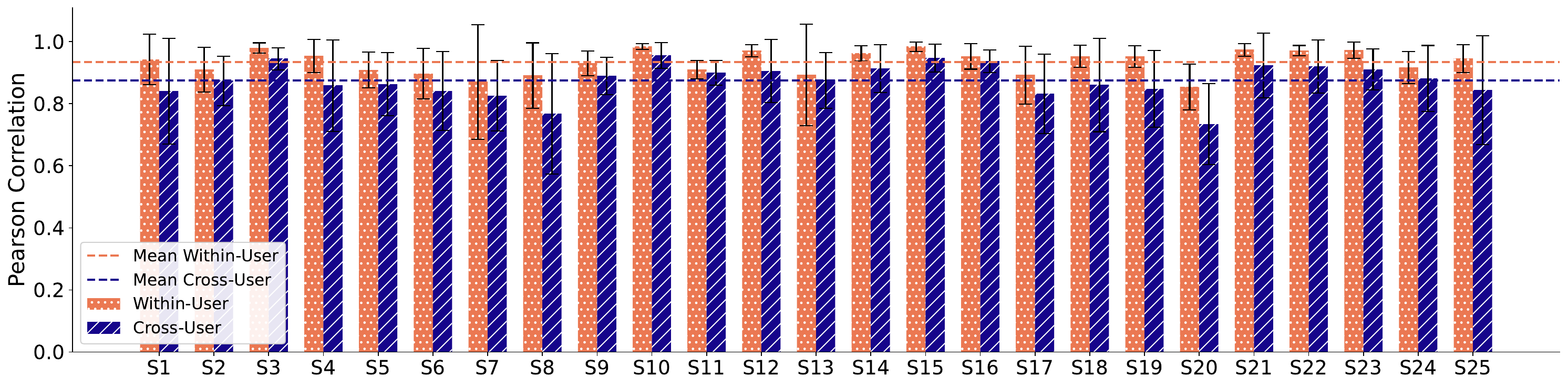}
  \caption{\red{\bf GCG similarity}}
\end{subfigure}

\caption{{\bf Waveform reconstruction performance across \red{25} subjects.} Results are shown for within-user and cross-user evaluations.}
\label{fig:within_cross}
\Description{This figure is composed of two bar charts, one above the other, that display the waveform reconstruction performance for two different signals across 25 subjects. The top chart, labeled (a) SCG similarity, shows the Pearson Correlation for the Seismocardiogram (SCG) signal. The bottom chart, labeled (b) GCG similarity, shows the Pearson Correlation for the Gyrocardiogram (GCG) signal. In both charts, each of the 25 subjects is represented by a pair of bars. The first bar in each pair, a stippled orange, represents the Within-User reconstruction performance, while the second bar, a solid navy blue, represents the Cross-User performance. Error bars are included to show standard deviation. Horizontal dashed lines on each chart indicate the Mean Within-User and Mean Cross-User correlation values.}
\end{figure*}

\begin{figure*}[htbp]
\centering
\includegraphics[keepaspectratio, width=.7\linewidth]
    {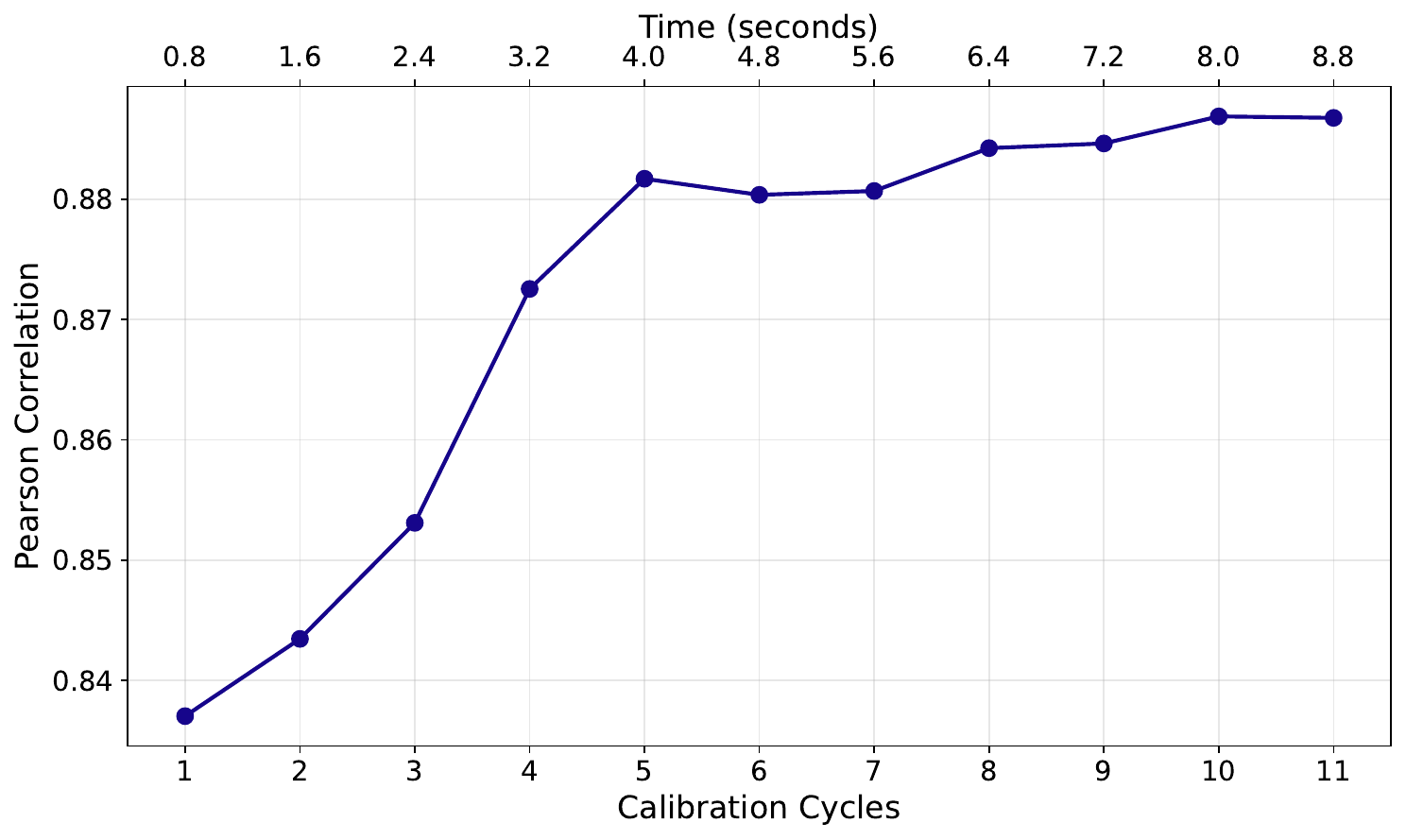}
  \caption{{\bf Effect of number of cardiac cycles used for calibration on cross-user SCG reconstruction performance.}}
  \label{fig:cal_cycle}
  \Description{This figure is a line graph illustrating the relationship between the number of calibration cycles and the performance of cross-user SCG reconstruction, as measured by the Pearson Correlation. The x-axis has two labels: the primary one at the bottom is "Calibration Cycles," ranging from 1 to 11, while a secondary one at the top is "Time (seconds)," ranging from 0.8 to 8.8 seconds. The y-axis, "Pearson Correlation," ranges from 0.82 to 0.89. The graph plots 11 data points connected by a dark blue line. The line shows a steep increase in Pearson Correlation from 1 to 4 calibration cycles, after which it stabilizes and plateaus. The specific data points are as follows: at 1 cycle (0.8 seconds), the Pearson correlation is 0.8370; at 2 cycles (1.6 seconds), it is 0.8435; at 3 cycles (2.4 seconds), it is 0.8531; at 4 cycles (3.2 seconds), it is 0.8725; at 5 cycles (4.0 seconds), it is 0.8817; at 6 cycles (4.8 seconds), it is 0.8804; at 7 cycles (5.6 seconds), it is 0.8807; at 8 cycles (6.4 seconds), it is 0.8843; at 9 cycles (7.2 seconds), it is 0.8846; at 10 cycles (8.0 seconds), it is 0.8869; and at 11 cycles (8.8 seconds), it is 0.8868. The standard deviations for each of these points are 0.1027, 0.1110, 0.0989, 0.0794, 0.0755, 0.0757, 0.0761, 0.0723, 0.0797, 0.0759, and 0.0771, respectively.}
\end{figure*}

\begin{figure*}
\centering
\includegraphics[keepaspectratio, width=\linewidth]
    {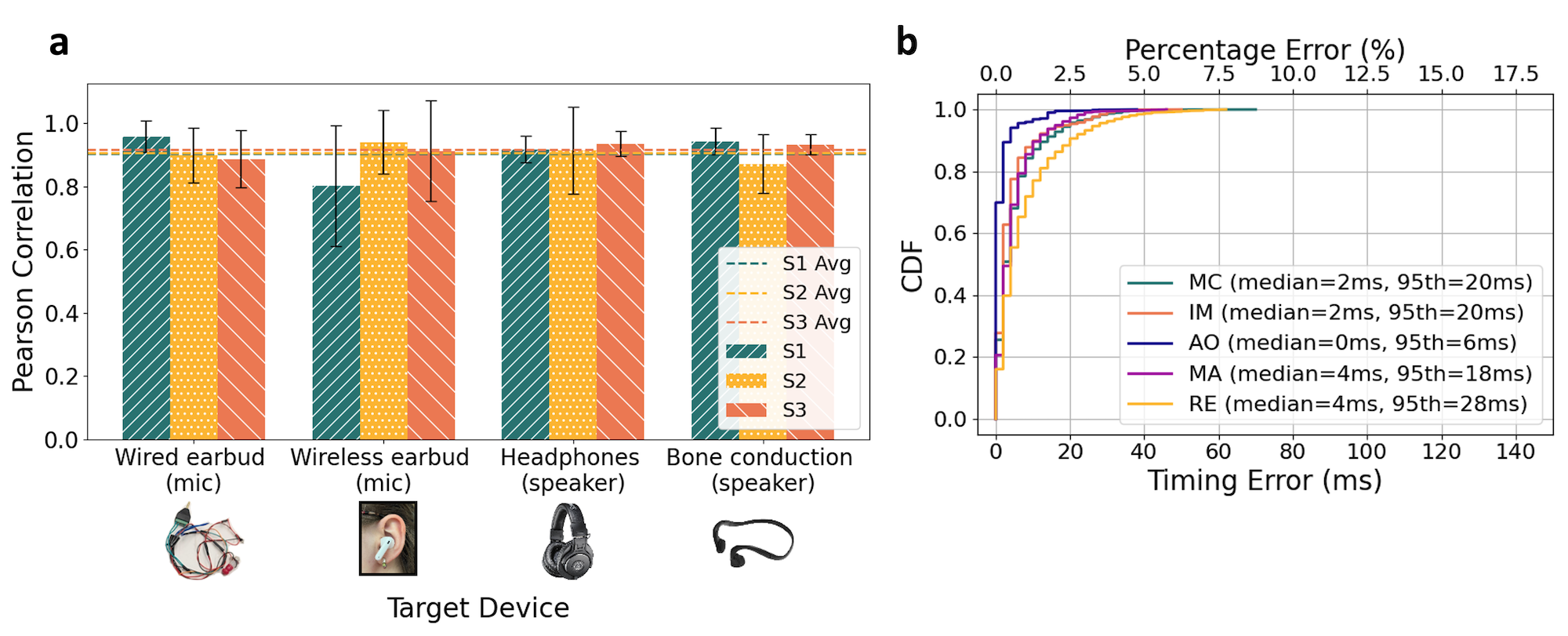}
    \vspace{-1em}
  \caption{{\bf Cross-device performance.} {\bf (a)} SCG waveform reconstruction {\bf (b)} fiducial point timing error. The ear-based heart sounds from different hearables are normalized to a single reference device (in-ear speaker of wired earbuds), and inference is performed on a model trained on that reference. Data is presented for three different subjects.}
  \label{fig:cross_device_main}
  \Description{The two sub-figures illustrate the cross-device performance of a system for cardiac signal processing. Sub-figure (a) is a bar chart showing SCG waveform reconstruction performance. The x-axis represents four Target Devices: a wired earbud (mic), a wireless earbud (mic), headphones (speaker), and a bone conduction headset (speaker), each with a corresponding image below. The y-axis is "Pearson Correlation," ranging from 0.0 to 1.0. For each device, three vertical bars show the results for three subjects: S1 (green), S2 (yellow), and S3 (orange). The Pearson correlations are as follows: for the wired earbud, S1 is 0.9578, S2 is 0.8994, and S3 is 0.887; for the wireless earbud, S1 is 0.802, S2 is 0.9408, and S3 is 0.9128; for the headphones, S1 is 0.9178, S2 is 0.913, and S3 is 0.9357; and for the bone conduction headset, S1 is 0.943, S2 is 0.8717, and S3 is 0.9329. Horizontal dashed lines represent the average similarity for each subject(S1, S2, S3) across all devices. Sub-figure (b) is a cumulative distribution function (CDF) plot showing fiducial point timing error. The bottom x-axis is "Timing Error (ms)" and the y-axis is "CDF." Five lines represent different methods: MC (median=2ms, 95th=20ms), IM (median=2ms, 95th=20ms), AO (median=0ms, 95th=6ms), MA (median=4ms, 95th=18ms), and RE (median=4ms, 95th=28ms).}
\end{figure*}



\noindent {\bf Within-user.} We first evaluate the reconstruction performance of SCG and GCG waveforms under a within-user setting, where both training and testing are performed on data from the same participant. When using the \textit{in-ear speaker of wired earbuds}, across \red{25} subjects, the reconstructed signals closely match the ground truth, achieving average Pearson correlation scores of \red{0.94 $\pm$ 0.04} for SCG and \red{0.93 $\pm$ 0.04} for GCG (Fig.~\ref{fig:within_cross}a). 

In comparison, when using the \textit{in-ear microphone of wireless earbuds}, the reconstruction performance was similar achieving 0.95 $\pm$ 0.05 for SCG and 0.95 $\pm$ 0.04 for GCG. These results demonstrate that we are able to achieve comparable reconstruction results from both in-ear speakers and microphones on two different devices.


\noindent {\bf Cross-user.} We then perform cross-user evaluation by training a model on all but one user, and fine-tuning it on five calibration cardiac cycles from the held-out user. Under this protocol, the system using the \textit{in-ear speaker} achieves mean Pearson correlation scores of \red{0.88 $\pm$ 0.07} for SCG and \red{0.88 $\pm$ 0.05} for GCG (Fig. \ref{fig:within_cross}b). Similarly, when using the \textit{in-ear microphone}, the system achieved scores of 0.84 $\pm$ 0.13 for SCG and 0.85 $\pm$ 0.09 for GCG.

We note that Pearson similarity scores of 0.90 is considered very high in the context of medical research~\cite{mukaka2012guide}. Prior work using mmWave sensors to reconstruct SCG achieved a correlation of 0.72~\cite{ha2020contactless}, while IMU-based earbuds reported a similarity of 0.92~\cite{fu2025enabling}. Our approach, based on ear-based speakers and microphones, provides a complementary modality that expands micro-cardiac monitoring across a wide range of hearables.

\noindent {\bf Effect of number of calibration cycles.} Fig.~\ref{fig:cal_cycle} shows how average Pearson correlation score varies with the number of calibration cycles. Performance increases from 83\% with a single cycle to 88\% with five cycles, after which performance gains plateau. Based on this trend, we select five calibration cycles for fine-tuning across all cross-user evaluations. Assuming an average adult heartbeat of 75~bpm, five calibration cycles corresponds to a brief 4-second calibration period.

\noindent {\bf Cross-device.} When a users buys a new hearable, we apply a zero-effort normalization strategy that removes the need for users to explicit perform a calibration step (see {\bf Sec.~\ref{sec:calib}}). Here, the in-ear speaker of wired earbuds served as the reference device, while the remaining four hearables (in-ear microphone of wired earbuds, in-ear microphone of wireless earbuds, microphone of over-ear headphones, or bone-conduction microphones) are target devices for testing.

We performed frequency-domain normalization on the ear-based cardiac sounds by using the mean of ten cardiac cycles between the reference and target device to compute equalization weights. Across data from four hearables and three users, the normalized waveforms achieved a mean Pearson correlation of 0.94 $\pm$ 0.02 when compared to the reference device.

We then trained a user-specific model for SCG reconstruction for each of the three users. Next, we applied the computed weights to produce normalized cardiac sounds data from the hearables. We then applied the models to the normalized waveforms, to obtain the reconstructed SCG waveforms. Across three users and four devices, the system achieved an average Pearson correlation of 0.91 $\pm$ 0.04 (Fig.~\ref{fig:cross_device_main}a), which is comparable to our within-user and cross-user results, demonstrating that our normalization approach enables cross-device generalization without explicit user calibration effort.

\subsection{Timing accuracy of micro-cardiac events}  
\label{sec:timing}

\begin{figure*}[htbp]
\centering
\begin{subfigure}{\linewidth}
  \centering
  \includegraphics[keepaspectratio, width=.49\linewidth]
    {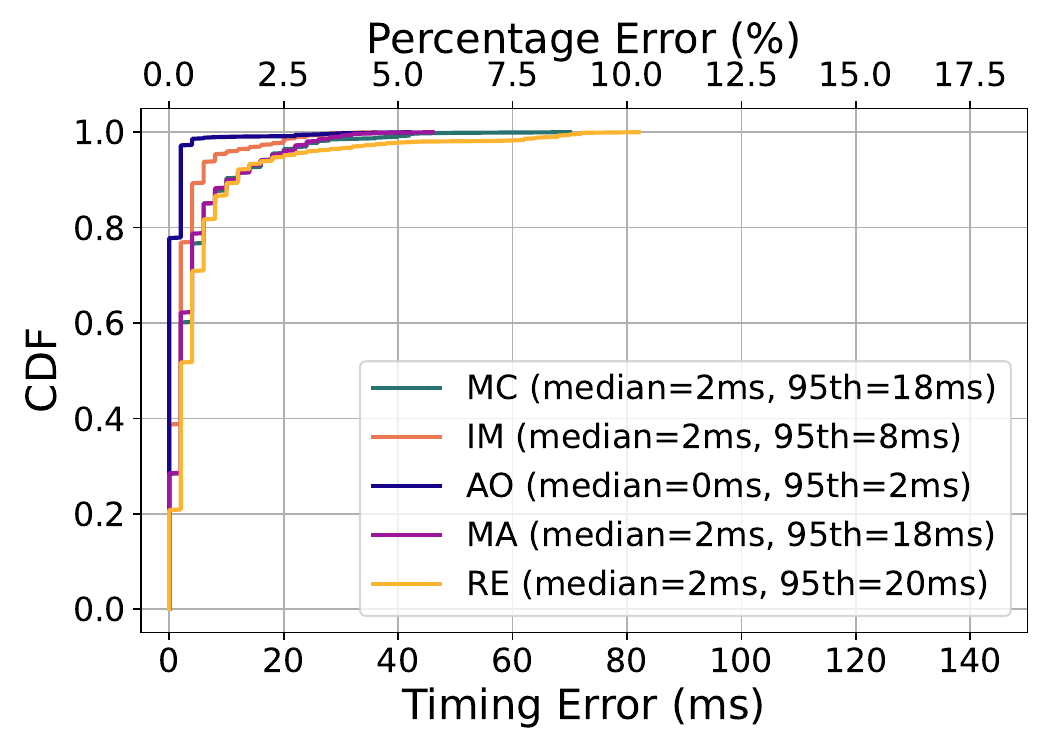}  
  \includegraphics[keepaspectratio, width=.49\linewidth]
    {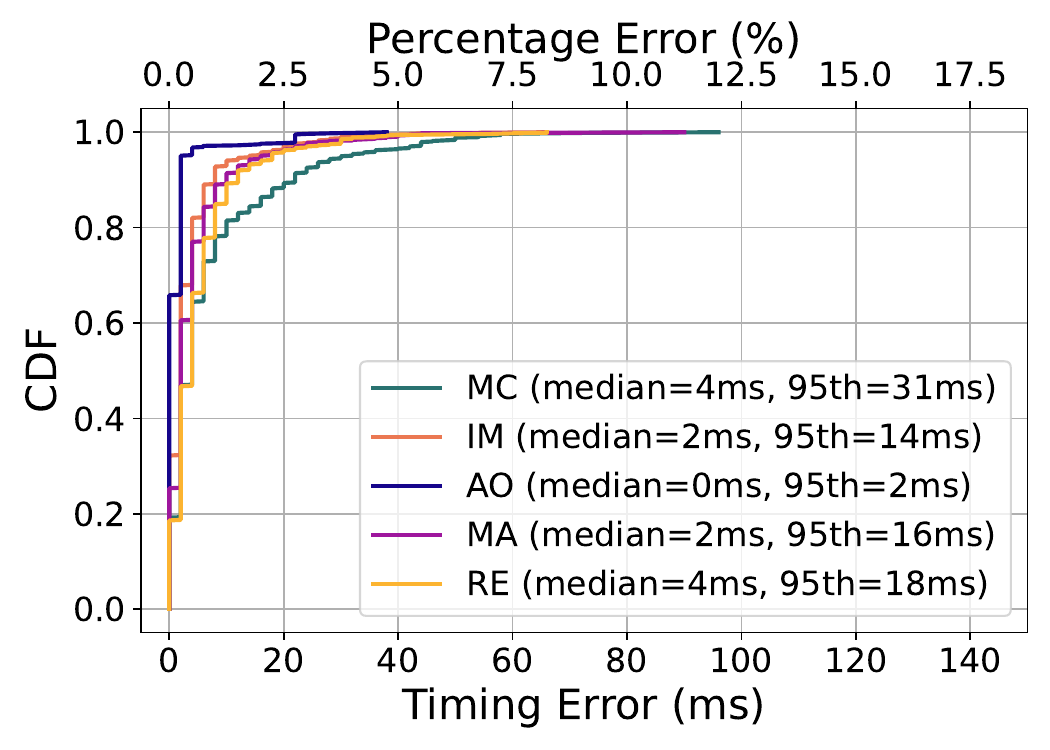}  
  \caption{\red{{\bf Within-user.} (Left) SCG, (Right) GCG}}
\end{subfigure}

\vspace{1em}

\begin{subfigure}{\linewidth}
  \centering
  \includegraphics[keepaspectratio, width=.49\linewidth]
    {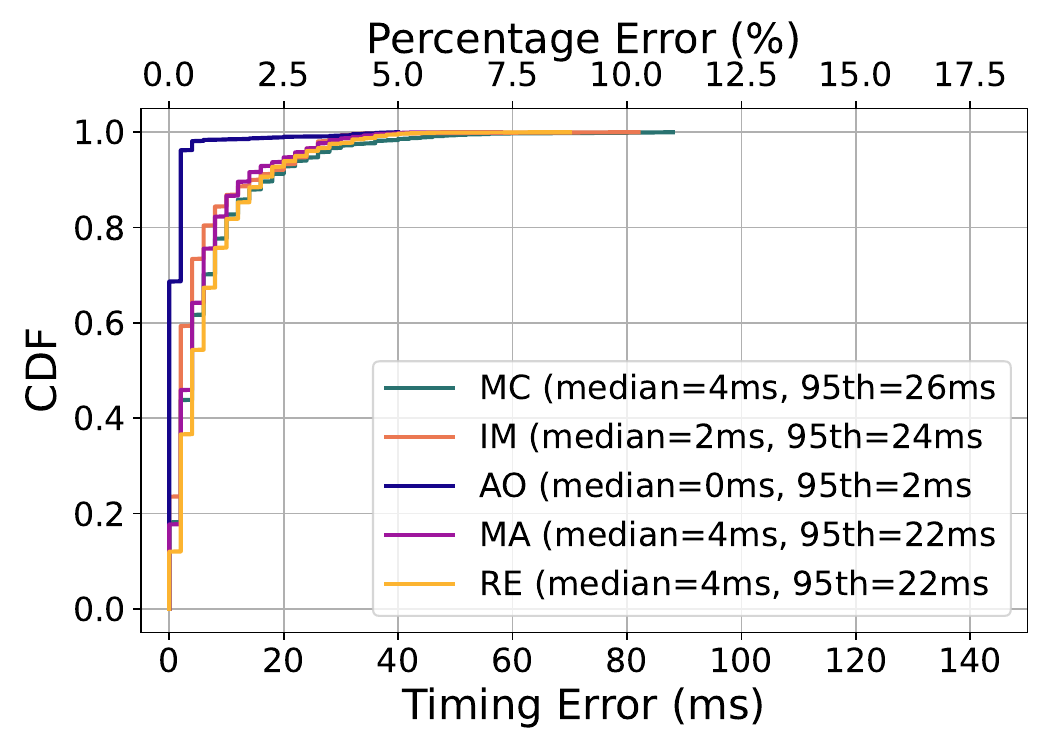}  
  \includegraphics[keepaspectratio, width=.49\linewidth]
    {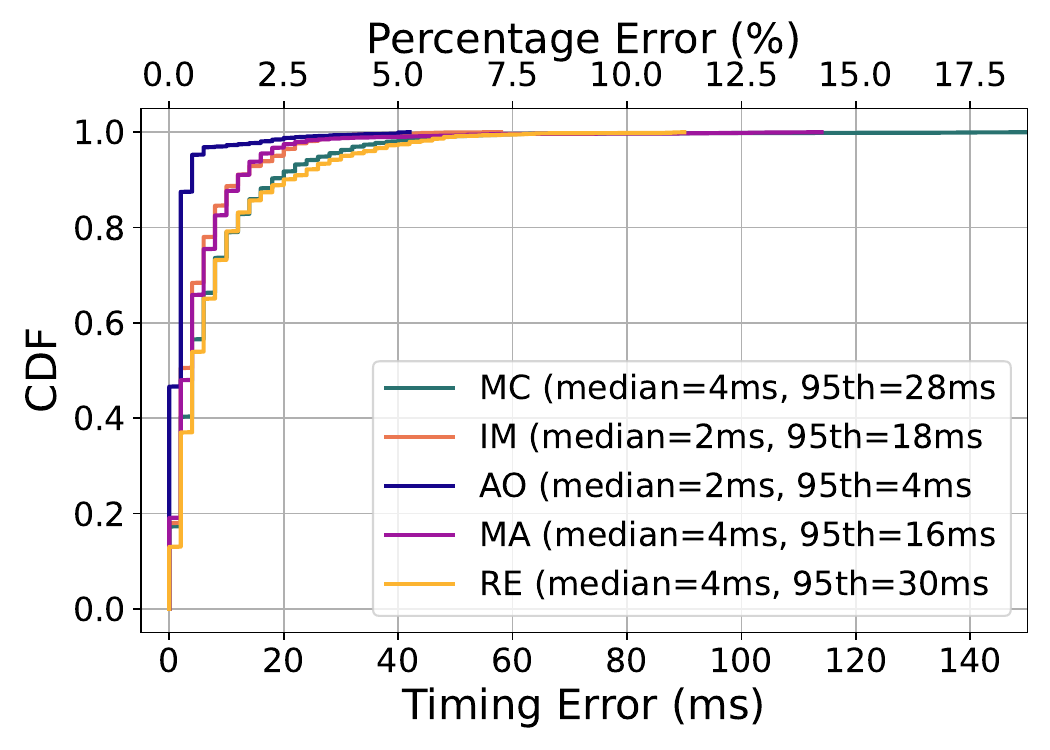}  
  \caption{\red{{\bf Cross-user.} (Left) SCG, (Right) GCG}}
\end{subfigure}

\caption{{\bf Timing accuracy of micro-cardiac events.} System performance is shown for SCG and GCG timing accuracy as collected on the in-ear speaker on earbuds.}
\label{fig:within_cross_cdf}
\Description{Four cumulative distribution function (CDF) plots showing timing accuracy of micro-cardiac events for SCG (seismocardiography) and GCG (gyrocardiography) measurements collected on in-ear speaker earbuds. The figure is organized in two rows: (a) Within-user performance and (b) Cross-user performance. Each row contains left panels showing SCG timing error and right panels showing GCG timing error.
In panel (a) within-user SCG (left), five overlapping curves represent different cardiac events: MC (mitral valve closure), IM (isovolumic moment), AO (aortic valve opening), MA (mitral valve anterior), and RE (rapid ejection). All curves rise steeply near 0ms timing error, with AO showing the best performance (median=0ms, 95th=2ms) and RE showing slightly lower performance (median=2ms, 95th=20ms). The within-user GCG panel (right) shows similar patterns with median timing errors between 0-4ms and 95th percentiles between 2-18ms.
In panel (b) cross-user evaluation, both SCG (left) and GCG (right) show slightly degraded but still strong performance. SCG median timing errors range from 0-4ms with 95th percentiles between 2-26ms, while GCG shows medians of 0-4ms with 95th percentiles between 4-30ms. All curves demonstrate that the system maintains sub-30ms timing accuracy at the 95th percentile across different cardiac event types in both within-user and cross-user scenarios, with AO and IM events showing the most precise timing detection.}
\end{figure*}

\begin{table}[t]
\centering
\setlength{\tabcolsep}{2.6pt}
\renewcommand{\arraystretch}{1.05}
\footnotesize


\resizebox{\linewidth}{!}{%
\begin{tabular}{lccccc}
\toprule
\multicolumn{6}{c}{\textbf{Within-user (SCG)}}\\
\midrule
 & \textbf{MC} & \textbf{IM} & \textbf{AO} & \textbf{MA} & \textbf{RE} \\
Median & 2.0[2.0,2.0] & 2.0[2.0,2.0] & 0.0[0.0,0.0] & 2.0[2.0,2.0] & 2.0[2.0,4.0] \\
95th  & 18.0[16.0,20.0] & 8.0[6.0,12.0] & 2.0[2.0,2.0] & 18.0[16.0,22.0] & 20.0[15.1,27.1] \\
\midrule
\multicolumn{6}{c}{\textbf{Within-user (GCG)}}\\
\midrule
Median & 4.0[2.0,4.0] & 2.0[2.0,2.0] & 0.0[0.0,0.0] & 2.0[2.0,2.0] & 4.0[2.0,4.0] \\
95th  & 30.6[26.0,37.2] & 14.0[10.0,20.0] & 2.0[2.0,4.0] & 16.0[14.0,20.0] & 18.0[16.0,20.6] \\
\bottomrule
\end{tabular}%
}

\vspace{4pt}

\resizebox{\linewidth}{!}{%
\begin{tabular}{lccccc}
\toprule
\multicolumn{6}{c}{\textbf{Cross-user (SCG)}}\\
\midrule
 & \textbf{MC} & \textbf{IM} & \textbf{AO} & \textbf{MA} & \textbf{RE} \\
Median & 4.0[4.0,4.0] & 2.0[2.0,2.0] & 0.0[0.0,0.0] & 4.0[4.0,4.0] & 4.0[4.0,4.0] \\
95th  & 26.0[24.0,26.0] & 24.0[22.0,24.0] & 2.0[2.0,2.0] & 22.0[20.0,22.0] & 22.0[22.0,24.0] \\
\midrule
\multicolumn{6}{c}{\textbf{Cross-user (GCG)}}\\
\midrule
Median & 4.0[4.0,4.0] & 2.0[2.0,4.0] & 2.0[2.0,2.0] & 4.0[4.0,4.0] & 4.0[4.0,4.0] \\
95th  & 28.0[26.0,30.0] & 18.0[18.0,20.0] & 4.0[4.0,6.0] & 16.0[16.0,18.0] & 30.0[29.8,34.0] \\
\bottomrule
\end{tabular}%
}

\vspace{4pt}

\resizebox{\linewidth}{!}{%
\begin{tabular}{lccccc}
\toprule
\multicolumn{6}{c}{\textbf{Cross-device (SCG)}}\\
\midrule
 & \textbf{MC} & \textbf{IM} & \textbf{AO} & \textbf{MA} & \textbf{RE} \\
Median & 2.0[2.0,4.0] & 2.0[2.0,2.0] & 0.0[0.0,0.0] & 4.0[2.0,4.0] & 4.0[4.0,4.0] \\
95th  & 20.0[18.0,22.0] & 20.0[16.0,22.5] & 6.0[4.5,8.0] & 18.0[16.0,18.0] & 28.0[26.0,30.0] \\
\bottomrule
\end{tabular}%
}
\caption{Timing error (ms) of micro-cardiac fiducial points between reconstructed and reference waveforms. Values are median and 95th percentile absolute error with 95\% confidence intervals.}
\label{tab:fiducial_timing}
\end{table}

We next evaluate the timing error of key micro-cardiac fiducial points (MC, IM, AO, MA, RE) between the reconstructed and reference waveforms. For each point, Table~\ref{tab:fiducial_timing} reports the median and 95th percentile absolute errors with 95\% confidence intervals (CI), along with a significance test across participants.

\noindent {\bf Within-user.} Fig.~\ref{fig:within_cross_cdf}a summarizes within-user performance, and Table~\ref{tab:fiducial_timing} reports the corresponding timing errors for SCG and GCG. Using a Wilcoxon signed-rank test across participants, we found no significant difference in the timing errors for any fiducial point (SCG: all $p>0.14$; GCG: all $p>0.38$). These values translate to relative median timing errors of 0.0--0.5\% and 95th percentile errors of 0.5--3.9\%, assuming an 800~ms cardiac cycle typical of a healthy adult heartbeat at 75~bpm.

\noindent {\bf Cross-user.} As illustrated in Fig.~\ref{fig:within_cross_cdf}b, Table~\ref{tab:fiducial_timing} reports cross-user timing errors for SCG and GCG. We again found no significant differences across participants (SCG: all $p>0.23$; GCG: all $p>0.31$). These correspond to relative median timing errors of 0.0--0.5\% and 95th percentile errors of 0.5--3.8\%.

We note that prior work~\cite{ha2020contactless} using mmWave for SCG reconstruction obtains similar median timing errors of 0.3--1.6\%. Furthermore, the within-session variability of fiducial point timing is often higher, ranging from 1--8~ms for SCG and GCG (Table~\ref{tab:variability2}).

\noindent {\bf Cross-device.} We evaluated the timing accuracy of reconstructed SCG fiducial points after applying cross-device normalization. As shown in Fig.~\ref{fig:cross_device_main}b and Table~\ref{tab:fiducial_timing}, the system achieved comparable timing accuracy to within-user and cross-user settings. Using a Wilcoxon signed-rank test across participants, we found no significant difference in timing errors for any fiducial point (all $p>0.16$). These correspond to relative median errors of 0.0--0.5\% and 95th percentile errors of 0.75--3.5\% of the cardiac cycle.



\subsection{Subgroup analysis} 
\begin{figure*}[htbp]
\centering
\includegraphics[keepaspectratio, width=\linewidth]
    {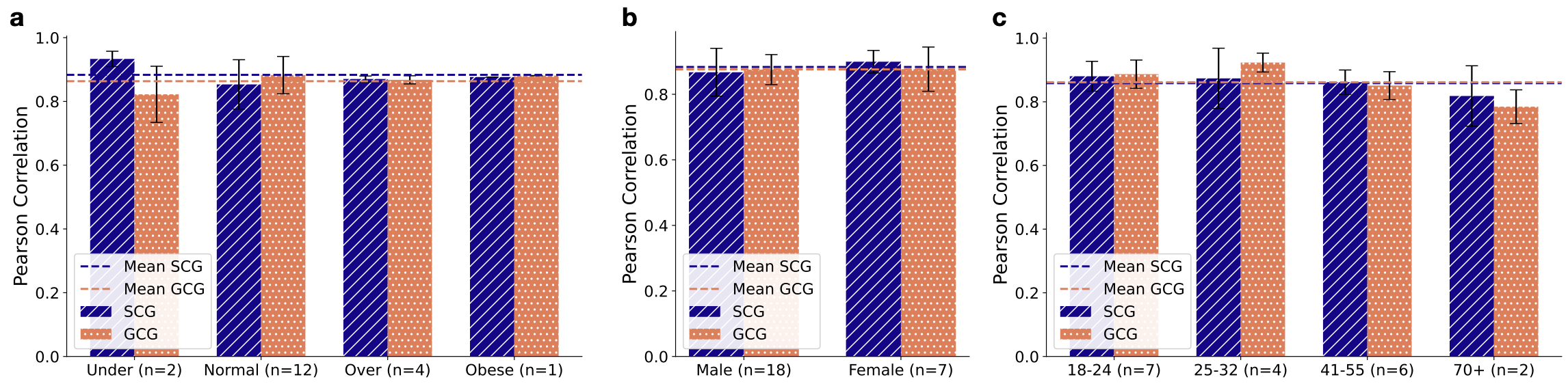}
  \caption{\red{{\bf Subgroup analysis on cross-user waveform reconstruction performance by} {\bf (a)} BMI, {\bf (b)} gender, {\bf (c)} age.}}
  \label{fig:subgroup}
  \Description{This figure contains two adjacent bar charts, each presenting a subgroup analysis of the cross-user waveform reconstruction performance. The y-axis for both charts is labeled Pearson Correlation, ranging from 0.0 to 1.0. The left chart displays performance based on body mass index (BMI) subgroups, while the right chart shows performance based on gender. In the left chart, the performance for the underweight subgroup (n=2) shows a mean Pearson correlation of 0.93 for SCG and 0.82 for GCG. The normal weight subgroup (n=12) shows a mean correlation of 0.85 with a standard deviation of 0.08 for SCG, and 0.88 with a standard deviation of 0.06 for GCG. The overweight subgroup (n=4) shows a mean correlation of 0.87 for SCG and 0.87 for GCG. Finally, the obese subgroup (n=1) has a mean correlation of 0.87 for SCG and 0.88 for GCG. In the right chart, the male subgroup (n=18) has a mean correlation of 0.86 with a standard deviation of 0.07 for SCG, and 0.88 with a standard deviation of 0.04 for GCG. The female subgroup (n=7) shows a mean correlation of 0.90 with a standard deviation of 0.03 for SCG, and 0.88 with a standard deviation of 0.07 for GCG. Error bars on all bars represent the standard deviation, and horizontal dashed lines indicate the overall mean Pearson correlation for SCG and GCG across all groups. For both figures, SCG refers to the color of dark blue, while GCG refers to the color of orange. The figure shows that among the 19 participants whom we have full demographic information in our study, 37\% (n=7) of them were among the age of 18-24, 21\% (n=4) were between 25-32, 32\% (n=6) were 41-55. and 11\% (n=2) were 70+. For the 18-24 age group, our system achieved average reconstruction correlation of 0.88 +/- 0.05 (SCG) and 0.89 +/- 0.04 (GCG). In the 25-32 age group, our system achieved reconstruction correlation of 0.87 +/- 0.09 (SCG) and 0.92 +/- 0.03 (GCG). In the 41-55 age group, our system achieved reconstruction correlation of 0.86 +/- 0.04 (SCG) and 0.85 +/- 0.04 (GCG). For the 70+ age group, our system achieved reconstruction correlation of 0.82 +/- 0.09 (SCG) and 0.78 +/- 0.05 (GCG). }
\end{figure*}

\noindent We performed a subgroup analysis on the cross-user waveform reconstruction results to evaluate system performance across different demographic dimensions (Fig.~\ref{fig:subgroup}).

\noindent {\bf Body mass index.} \red{Among the 25 participants that participated in our study, only 19 of them shared their BMI data. }\red{11\% ($n=2$)} of subjects were underweight ($BMI < 18.5$), \red{63\% ($n=12$)} had normal weight ($18.5 \leq BMI < 24.9$), \red{21\% ($n=4$)} were overweight ($25.0 \leq BMI < 29.9$), and \red{5\% ($n=1$)} were obese ($BMI \geq 30$). For the underweight group, the system achieved average reconstruction correlation of \red{$0.93 \pm 0.02$} (SCG) and \red{$0.82 \pm 0.09$} (GCG). In the normal weight group, similarities were \red{$0.85 \pm 0.08$} (SCG) and \red{$0.88 \pm 0.25$} (GCG). Among overweight subjects, results were \red{$0.87 \pm 0.09$} (SCG) and \red{$0.87 \pm 0.01$} (GCG). Finally, in the obese group, the system reached $0.90 \pm 0.09$ (SCG) and $0.89 \pm 0.10$ (GCG).

\noindent {\bf Sex.} In our study, 28\% \red{($n=7$)} of subjects were female and 72\% \red{($n=18$)} of subjects were male. When comparing system performance, female subjects had an SCG reconstruction similarity of \red{$0.90 \pm 0.03$} and a GCG similarity of \red{$0.88 \pm 0.07$}, while male subjects had an SCG similarity of \red{$0.87 \pm 0.07$} and a GCG similarity of \red{$0.88 \pm 0.05$}.

\noindent \red{{\bf Age.} Among the 19 participants whom we have full demographic information in our study, 37\% ($n=7$) of them were among the age of 18-24, 21\% ($n=4$) were between 25-32, 32\% ($n=6$) were 41-55. and 11\% ($n=2$) were 70+. For the 18-24 age group, our system achieved average reconstruction correlation of $0.88 \pm 0.05$ (SCG) and $0.89 \pm 0.04$ (GCG). In the 25-32 age group, our system achieved reconstruction correlation of $0.87 \pm 0.09$ (SCG) and $0.92 \pm 0.03$ (GCG). In the 41-55 age group, our system achieved reconstruction correlation of $0.86 \pm 0.04$ (SCG) and $0.85 \pm 0.04$ (GCG). For the 70+ age group, our system achieved reconstruction correlation of $0.82 \pm 0.09$ (SCG) and $0.78 \pm 0.05$ (GCG). The results show that our system continues to achieve reliable performance in older adults.
}

\noindent {\bf Demographic classification.} We perform an exploratory analysis to assess whether the reconstructed SCG signals could be used to infer demographic attributes, and focus on the task of sex classification. From the reconstructed signals, we extracted features including linear predictive coding coefficients, mean and standard deviation. Using signals from the in-ear speaker of earphones, the binary classification accuracy reached 89\%. We did not perform a similar analysis for BMI, as our dataset does not have a sufficiently balanced distribution across BMI categories. Our classification result aligns with prior work~\cite{tokmak2023investigating} which report that sex-related differences in body composition influence SCG characteristics and can be used for classification. 

\subsection{\red{Generalization across real-world conditions}}

\begin{figure*}[htbp]
\centering
\includegraphics[keepaspectratio, width=\linewidth]
    {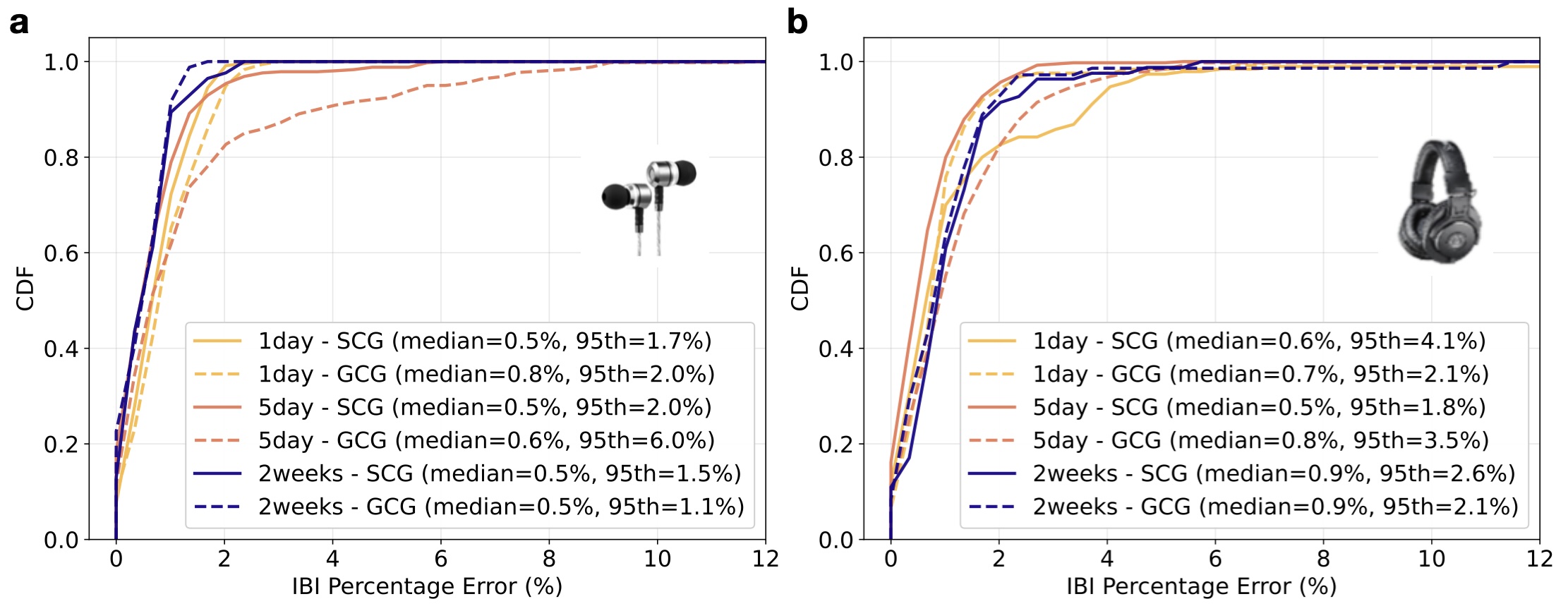}
    \caption{\red{{\bf Longitudinal effect of time on system performance.} CDFs show the system's IBI percentage error for SCG and GCG reconstruction across one-day, five-day, and two-week intervals for both the {\bf (a)} wired earphones and {\bf (b)} headphones. Performance remains stable over time, demonstrating that the model generalizes across natural day-to-day variations in heart rate.}}
  \label{fig:longitudinal}
  \Description{ The figure contains two side by side cumulative distribution plots of inter beat interval, IBI, percentage error, with the x axis ranging from 0 to 12 percent error and the y axis showing CDF from 0 to 1. Panel a, wired earphones, and panel b, headphones, each plot six curves comparing SCG and GCG reconstruction over three time gaps, 1 day, 5 days, and 2 weeks, with solid lines for SCG and dashed lines for GCG. In both panels, all curves rise steeply and reach about 0.9 to 1.0 CDF by roughly 2 to 3 percent error, indicating most IBI errors are small and performance is similar across time gaps. Reported medians and 95th percentiles are: panel a, 1 day SCG 0.5 and 1.7 percent, 1 day GCG 0.8 and 2.0 percent, 5 day SCG 0.5 and 2.0 percent, 5 day GCG 0.6 and 6.0 percent, 2 weeks SCG 0.5 and 1.5 percent, 2 weeks GCG 0.5 and 1.1 percent. Panel b, 1 day SCG 0.6 and 4.1 percent, 1 day GCG 0.7 and 2.1 percent, 5 day SCG 0.5 and 1.8 percent, 5 day GCG 0.8 and 3.5 percent, 2 weeks SCG 0.9 and 2.6 percent, 2 weeks GCG 0.9 and 2.1 percent. Small inset images depict the corresponding earphones in panel a and headphones in panel b.}
\end{figure*}

\begin{figure*}[htbp]
\centering
\includegraphics[keepaspectratio, width=\linewidth]
    {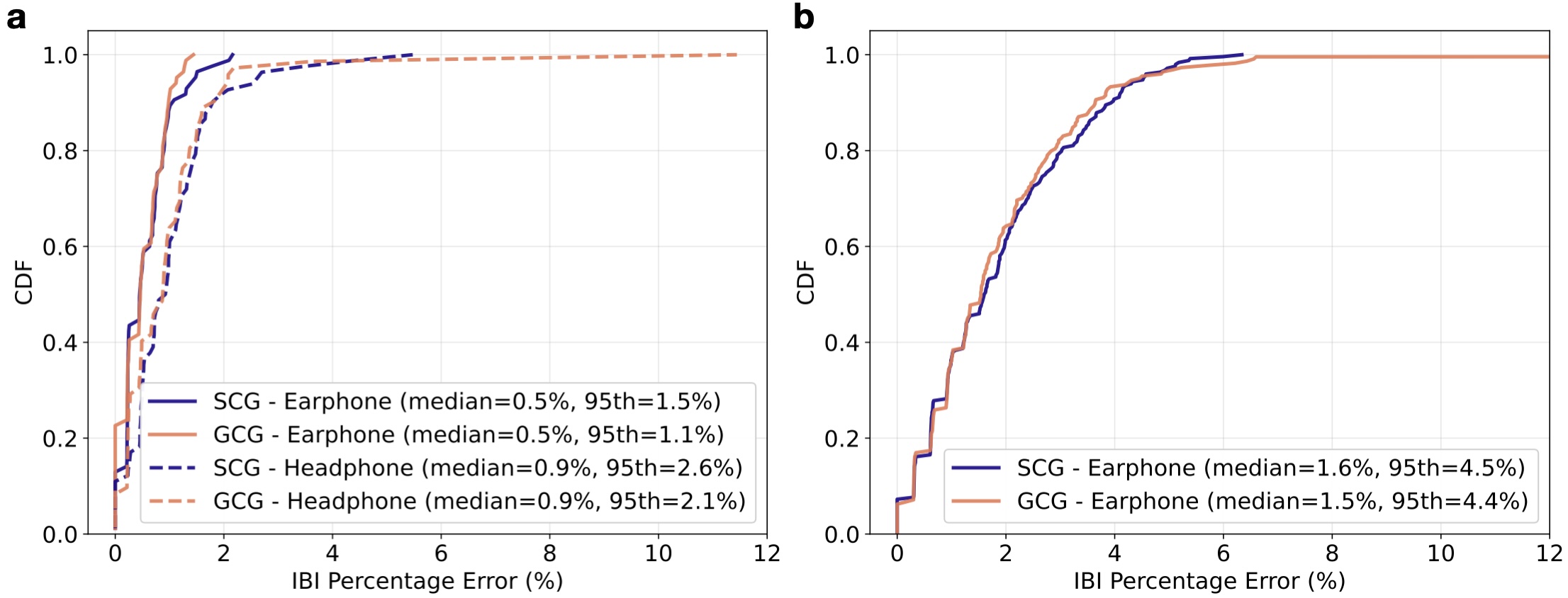}
    \caption{\red{{\bf Generalization across scenarios that perturb heartrate: (a)} sleeping and {\bf (b)} after exercise. CDFs illustrate the system’s IBI percentage for reduced heart rate during a nap and elevated heart rate following exercise. Results show that the model maintains accurate inter-beat timing across both reduced and increased heart rate conditions.}}
  \label{fig:exercise_and_nap}
  \Description{Two side-by-side cumulative distribution function (CDF) plots showing the generalization of a heart rate monitoring system across different scenarios. Panel (a) shows IBI percentage error during sleeping with reduced heart rate, displaying four curves for SCG-Earphone, GCG-Earphone, SCG-Headphone, and GCG-Headphone configurations. All curves rise steeply near 0\% error, with median errors between 0.5-0.9\% and 95th percentiles between 1.1-2.6\%. Panel (b) shows IBI percentage error after exercise with elevated heart rate, where the curves rise more gradually, with median errors around 1.5-1.6\% and 95th percentiles around 4.4-4.5\%. Both panels demonstrate that the system maintains accurate inter-beat interval timing across both reduced and increased heart rate conditions, with performance remaining consistent between SCG and GCG methods using either earphone or headphone form factors.
}
\end{figure*}

\red{Heart rate fluctuates naturally across days, increases after physical activity, and decreases during sleep. To evaluate whether our system can generalize across these real-world variations, we conducted three sets of experiments on two participants: (1) longitudinal testing across multiple days and weeks, (2) evaluation while lying down during a nap, and (3) evaluation after exercise.}

\red{We first trained the model on each user’s calm, static data using our multi-cycle training strategy and stretching-based augmentation. We then tested the model on data collected on different days, during sleep-like conditions, and after exercise. To quantify inter-beat accuracy, we report the inter-beat interval (IBI) percentage error, defined as the absolute timing error normalized by the true inter-beat interval:}

\red{
\begin{equation}
    \mathrm{IBI}_{\text{err}} = 100 \times 
    \frac{\left| \mathrm{IBI}_{\text{pred}} - \mathrm{IBI}_{\text{true}} \right|}
         {\mathrm{IBI}_{\text{true}}}
\end{equation}
}

\noindent \red{{\bf Longitudinal performance across time.} Our system maintains consistently high performance across 1-day, 5-day, and 2-week spans for both earphones and headphones (Fig.~\ref{fig:longitudinal}).
For earphones, SCG IBI percentage errors remained stable over time, with median errors of 0.5\%, 0.5\%, and 0.5\% at 1 day, 5 days, and 2 weeks, respectively, and corresponding 95th-percentile errors of 1.7\%, 2.0\%, and 1.5\%. For GCG, median errors were 0.8\%, 0.6\%, and 0.5\%, with 95th-percentile errors of 2.0\%, 6.0\%, and 1.1\%. For headphones, performance was similarly stable, with SCG median errors of 0.6\%, 0.5\%, and 0.9\% at 1 day, 5 days, and 2 weeks, and 95th-percentile errors of 4.1\%, 1.8\%, and 2.6\%. GCG median errors were 0.7\%, 0.8\%, and 0.9\%, with 95th-percentile errors of 2.1\%, 3.5\%, and 2.1\%. These results demonstrate that once trained on calm data, the model generalizes across natural day-to-day and week-to-week heart rate changes.}


\noindent \red{{\bf Lying down while sleeping.} Here we had the participants take a one hour nap, during which their heart rate slowed, under these conditions our system continued to perform robustly (Fig.~\ref{fig:exercise_and_nap}a). When using earphones, SCG median error was 0.5\% (95th percentile 1.5\%), and GCG median error was 0.5\% (95th percentile 1.1\%). For headphones, SCG had a median of 0.9\% (95th percentile 2.6\%), and GCG a median of 0.9\% (95th percentile 2.1\%). Performance remains in line with results measured when the user is awake, indicating robustness to naturally reduced heart rates during rest and sleep-like conditions.}


\noindent \red{{\bf Exercise.} We had participants jog on the spot for 10 minutes and measured system performance after this exercise session (Fig.~\ref{fig:exercise_and_nap}b). For earphones, SCG median errors were 1.6\% (95th percentile 4.5\%), and GCG median error increased to 1.5\% (95th percentile 4.4\%). These errors are modest given the rapid heart rate changes post-exercise.}


\subsection{Motion artifact removal}
\begin{figure*}
  \centering
  \includegraphics[width=.8\linewidth]{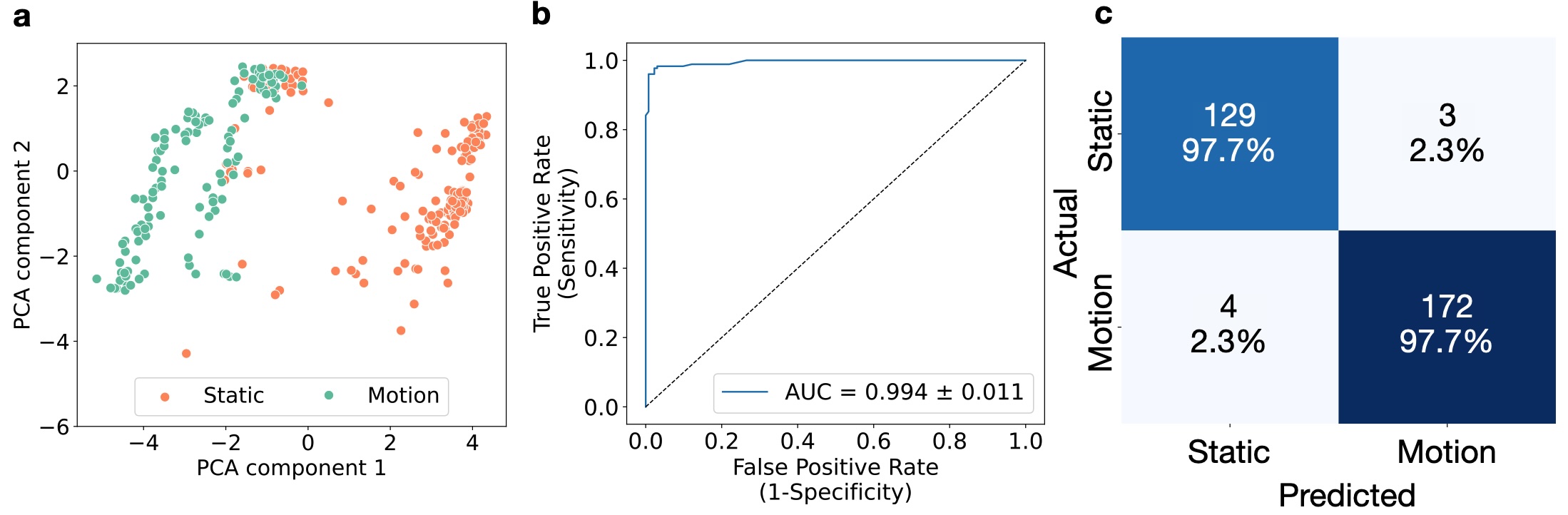}
  \vspace{-1em}
  \caption{{\bf Motion artifact removal pipeline performance.} Evaluation is performed on in-ear audio segments when the user is static and moving. {\bf (a)} PCA projection of the first two components of MFCC audio features illustrating separability between the two classes. {\bf (b)} ROC curve from five-fold cross validation. {\bf (c)} Confusion matrix indicating optimal operating point on the ROC curve.}
  \label{fig:motion}
  \Description{The three sub-figures illustrate the system's performance in distinguishing between static and motion states. Sub-figure 'a', a scatter plot, displays the results of a Principal Component Analysis (PCA) projection of audio features. The x-axis is labeled "PCA component 1" and the y-axis is labeled "PCA component 2". The data points are grouped into two distinct clusters: one cluster, represented by orange dots, corresponds to the 'Static' class, and the other, represented by green dots, corresponds to the 'Motion' class. The clear visual separation between these two clusters indicates that the chosen features are effective for distinguishing between static and motion conditions. Sub-figure 'b' is a Receiver Operating Characteristic (ROC) curve, which shows the trade-off between the True Positive Rate (Sensitivity) on the y-axis and the False Positive Rate (1-Specificity) on the x-axis. The curve rises steeply and closely follows the top-left corner of the graph, which signifies high classification performance. The area under the curve (AUC) is given as 0.994 with a standard deviation of 0.011, further quantifying the system's excellent performance. Sub-figure 'c' is a confusion matrix that provides a detailed breakdown of the classification results. The rows represent the actual classes, 'Static' and 'Motion', and the columns represent the predicted classes. For the 'Static' class, 129 instances were correctly predicted as 'Static' (97.7\%), while 3 were incorrectly predicted as 'Motion' (2.3\%). For the 'Motion' class, 172 instances were correctly predicted as 'Motion' (97.7\%), and 4 were incorrectly predicted as 'Static' (2.3\%). Overall, this figure effectively demonstrates the high accuracy and robust performance of the motion artifact detection system.}
\end{figure*}

We evaluate the performance of our motion artifact removal pipeline on an in-ear microphone audio dataset from the wireless earbuds (Fig.~\ref{fig:waveforms}a) collected when the user is static and moving (brow raising, chewing, nodding, talking, and walking), which we describe in {\bf Sec.~\ref{sec:motion_system}}. To assess the separability between these two classes, we applied principal component analysis (PCA) on the MFCCs of the audio and visualized the projection of the first two components (Fig.~\ref{fig:motion}a) which shows separability between the classes.

We trained a Random Forest classifier to distinguish between the static and motion recordings using 10-second segments as input. The model was evaluated using 5-fold cross-validation, achieving a ROC AUC of 0.994 $\pm$ 0.011 (Fig.~\ref{fig:motion}b) and a mean accuracy of 97.7 $\pm$ 2.2\% (Fig.~\ref{fig:motion}c).

{\color{red}

\begin{figure*}[h]
\centering
\includegraphics[keepaspectratio, width=\linewidth]
    {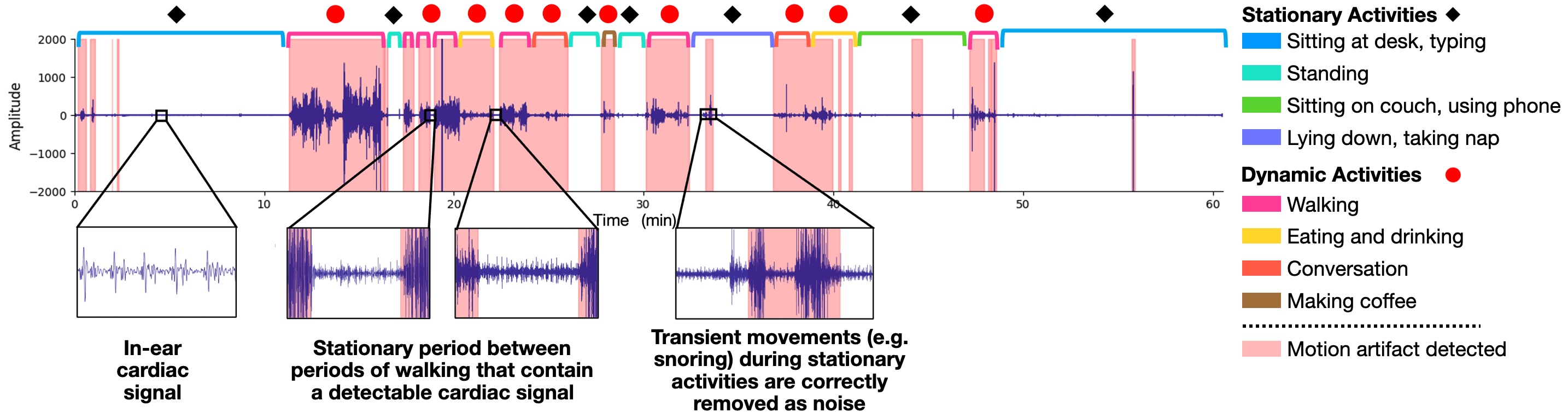}
\caption{\red{{\bf Free-living deployment across different stationary and dynamic activities.} The figure shows the in-ear audio signal from the speakers of wired earbuds with activity labels. Red shaded segments denote periods flagged by our motion detection algorithm. All dynamic activities are correctly flagged, no stationary activities were flagged, and the cardiac signal is visible during these stationary periods.}}
\label{fig:onehour}
    \Description{ This figure includes one-hour of real-world recording from a participant wearing wireless earbuds: It is a one-hour time-domain signal with activity labels, showing detected motion artifacts in red mask. Static activities include working at desk, standing, sitting on couch and using phone, and taking a nap while lying down. Dynamic activities include walking, eating and drinking, having a conversation, and making coffee. Motion activities and static activities happen intermittently, representing a typical one-hour time of a person's workday life. Importantly, our system captures the in-ear cardiac signals during brief stationary intervals between dynamic activities, and even short pauses between walking are sufficient to extract a clean cardiac signal. } 
\end{figure*}

\noindent \red{{\bf Free-living deployment.} To evaluate our motion artifact removal pipeline in realistic conditions, we conducted a one-hour recording session in which a participant wore the \textit{in-ear speaker of wired earbuds} while going about daily activities. As shown in Fig.~\ref{fig:onehour}, the system accurately detects and discards all motion-induced artifacts throughout the recording. In contrast, during stationary activities such as desk work our system consistently captures clean ear-based cardiac signals. Importantly, our system captures the in-ear cardiac signals during brief stationary intervals between dynamic activities, and even short pauses between walking are sufficient to extract a clean cardiac signal. Finally, the system is also able to correctly suppress transient noise that occurs during stationary activities such as snoring when the participant is taking a nap. Overall, this real-world evaluation demonstrates that our motion removal pipeline can reliably identify clean cardiac segments across a wide range of everyday scenarios.}




}

\subsection{Benchmarks}
\label{sec:benchmark}

\noindent {\bf Effect of device remounting.} To evaluate the effect of device-remounting on end-to-end system performance, we first trained a within-session model between paired hearable and smartphone recordings as a baseline. We then trained our model on remounted recordings, where both the hearable and the smartphone were independently removed and remounted, and tested on unseen remounted data. The baseline model achieved an average similarity score of 0.95 ± 0.02, while models evaluated on unseen remounted data achieved an average similarity score of 0.89 ± 0.02. To put this in context, the intrinsic variability of the SCG waveform within a single user is 0.86 (Table~\ref{tab:variability1}). \textit{This suggests that device-remounting has only a modest effect on end-to-end system performance.}

\begin{figure}[h]
\centering
\includegraphics[keepaspectratio, width=\linewidth]
    {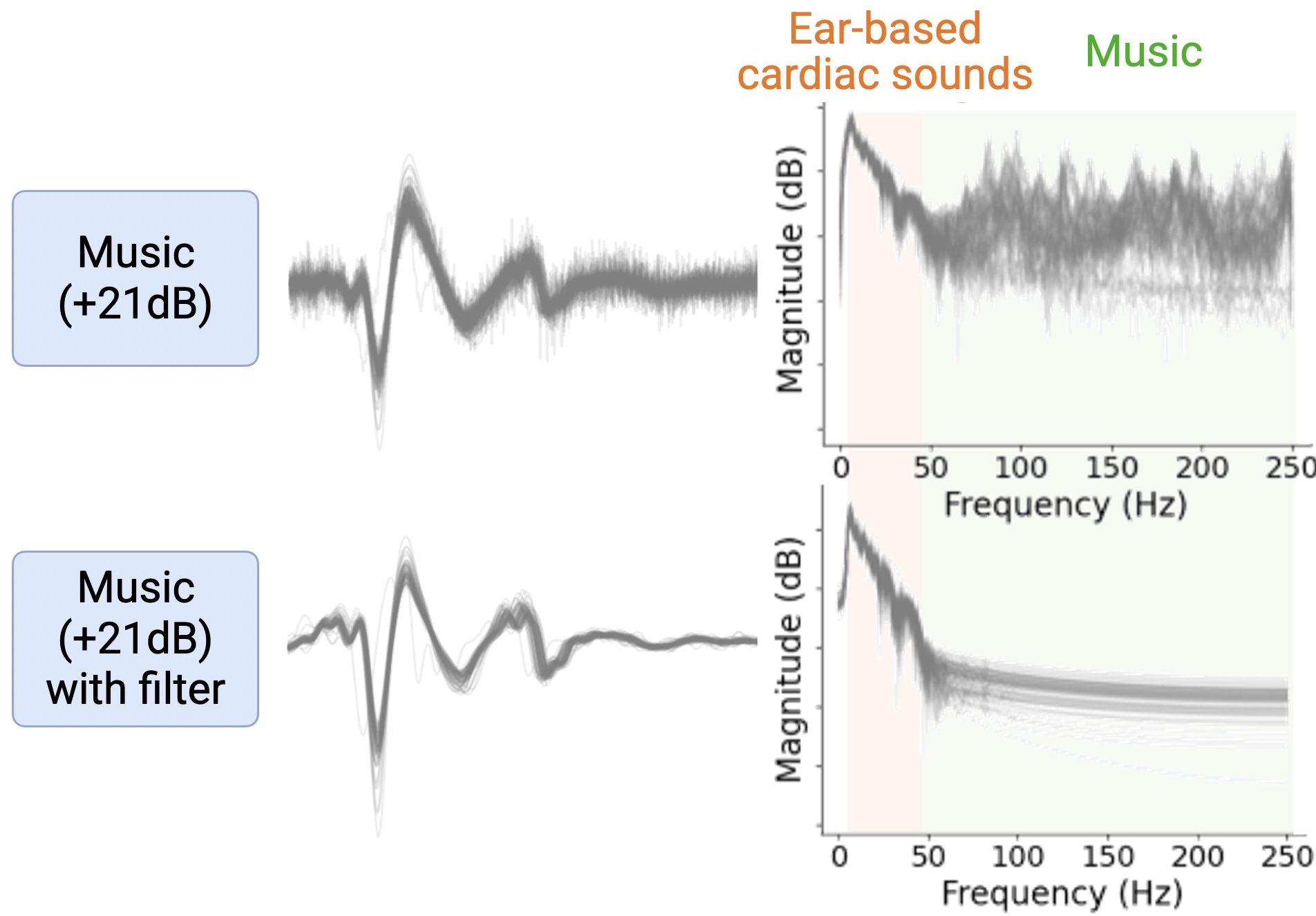}
    \vspace{-1em}
  \caption{{\bf Effect of music playback.} As micro-cardiac signals and music largely occupy non-overlapping frequency bands, the overall shape of the cardiac cycle and fiducial points are visible even without a 5--45~Hz bandpass filter.}
  \label{fig:music}
    \Description{This figure is composed of two rows of plots, demonstrating the effect of a frequency filter on reducing music interference in an ear-based cardiac sound signal. The top row of plots illustrates the raw, unfiltered signal. The top-left plot is a time-domain view of the cardiac signal, which is heavily obscured by high-amplitude, noisy waveforms caused by music played at +21 dB. The top-right plot shows the corresponding frequency-domain representation of this signal. It shows high-magnitude values across the entire frequency range from 0 Hz to 250 Hz, with a gradual roll-off, indicating the strong presence of broadband music noise. A vertical orange bar highlights the 0 to 45 Hz range, labeled "Ear-based cardiac sounds," while the rest of the plot to 250 Hz is labeled "Music" in a green font and in green light green background. The bottom row of plots shows the same signal after it has been processed with a frequency filter. The bottom-left plot, a time-domain view, now shows a much cleaner signal. The distinct, periodic cardiac waveforms are clearly visible, and the overall signal noise is significantly reduced. The bottom-right plot, the frequency-domain representation, shows that the filter has effectively attenuated the high-frequency content, with the signal magnitude dropping sharply after 45 Hz. The low-frequency content (from 0 to 45 Hz) is preserved, as shown by the orange-highlighted area, while the music frequencies above 45 Hz are significantly suppressed and in light-green-highlighted area. This figure effectively highlights how a simple frequency filter can isolate the low-frequency cardiac sound signals from high-frequency music noise.}
\end{figure}

\noindent {\bf Effect of music playback.} We evaluated the effect of music playback by having a subject wear wireless earbuds while music was played through the in-ear speaker at volume levels of +6, +9, +12, +15, +18, and +21~dB above ambient noise, with recordings collected from the in-ear microphone.

As shown in Fig.~\ref{fig:music}, the energy from music and the SCG signal largely reside in non-overlapping frequency ranges with music being concentrated in higher frequencies ($>$45~Hz). As a result, the overall shape of the SCG waveform and the fiducial points can still be observed both with and without a 5--45~Hz band-pass filter in the time domain. We evaluated end-to-end system performance by assessing waveform-reconstruction similarity for the different sound levels. Across all sound levels, waveform similarity remains high, with an average Pearson correlation of $0.95 \pm 0.01$ with frequency filtering and $0.94 \pm 0.01$ without. \textit{This shows that our system continues to perform even in the presence of music playback.}


\begin{figure}[h]
\centering
\includegraphics[keepaspectratio, width=\linewidth]
    {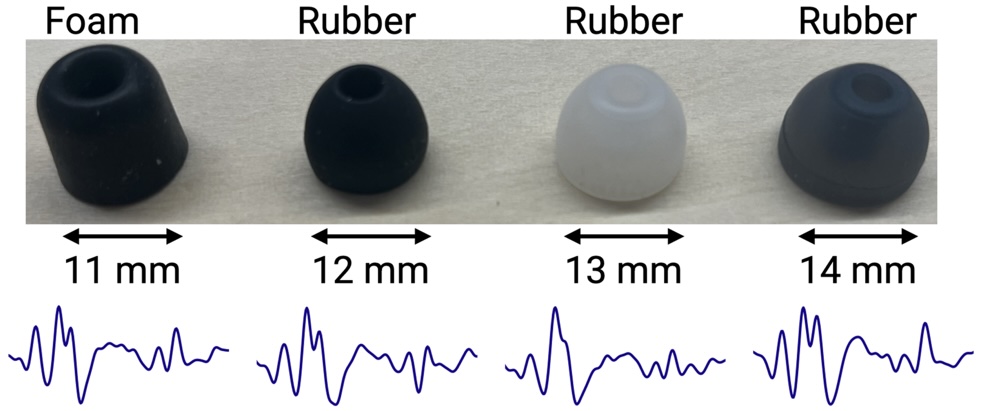}
    \vspace{-1em}
\caption{{\bf Effect of ear tip on ear-based cardiac sounds.} Mean cardiac cycles recorded from one participant using four different ear-tip types.}  \label{fig:eartips}
    \Description{This figure illustrates the effect of ear tips size. It is a composite image showing four vertical columns. Each column contains a photograph of an earbud tip at the top and a corresponding line graph of a signal waveform at the bottom. From left to right, the first column displays a black foam ear tip measured at 11 millimeters. The second column shows a black rubber ear tip measured at 12 millimeters. The third column contains a white rubber ear tip measured at 13 millimeters. Finally, the fourth column shows a gray rubber ear tip, the largest at 14 millimeters. For each waveform figure, it shows the mean cardiac cycle recorded from one participant using each ear-tip types.}
\end{figure}

\noindent {\bf Effect of ear tip diameter and material.} We selected one foam ear tip and three rubber ear tips with diameters of 11, 12, 13, and 14~mm, that were coupled to the in-ear earphone (Device 3 in Fig.~\ref{fig:wearing_devices}b). A single participant wore the device with each ear tip, and ear-based cardiac signals were recorded from the built-in speaker. Fig.~\ref{fig:eartips} shows the waveforms recorded from each of the ear tips. The average Pearson correlation across the four eartips was $0.88 \pm 0.06$, suggesting that for the common ear tip sizes and materials used for adults, it has minimal effect on the morphology of the measured cardiac signals.  

To further examine whether ear tips influence system performance, we evaluated cross–ear tip reconstruction accuracy. Specifically, we trained the model using data recorded with one ear tip and tested it on signals recorded with the others. Reconstruction similarity remained consistently high across ear tips, with an average Pearson correlation of $0.89 \pm 0.05$. These results indicate that variations in ear tip size or material have little impact on overall system performance.

\noindent {\bf System latency.} We evaluated the inference latency of our autoencoder model on a HUAWEI Pura 70 Ultra smartphone by exporting our PyTorch model to the ONNX format for mobile deployment. The system was run on in-ear audio chunks of 10000~ms, yielding a mean and standard deviation inference latency for each segment of 19.4 $\pm$ 2.4~ms, satisfying the timing requirements for real-time execution.

\subsection{User experience survey}

\red{Beyond evaluating waveform reconstruction performance, we sought to understand aspects of the user experience that influence real-world adoption rates and long-term engagement in everyday contexts. Specifically, we seek to understand user comfort, perceived unobtrusiveness, and attitudes towards in-ear cardiac sensing.}

\begin{figure*}[h]
\centering
  \includegraphics[keepaspectratio, width=\linewidth]{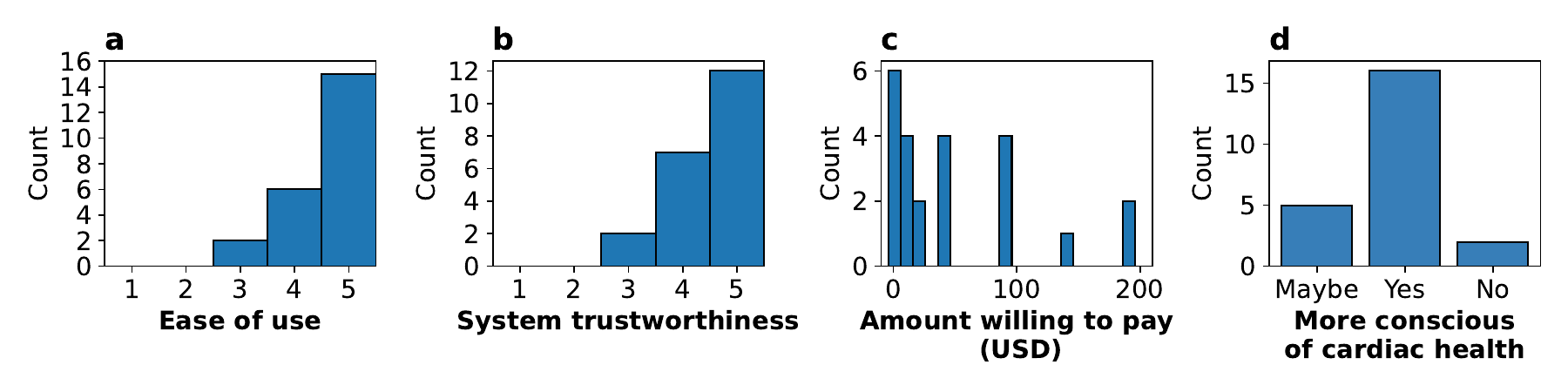}
    \vspace{-2.5em}
  \caption{{\bf User experience survey \red{($n=23$)}.} Histograms summarize user perceptions across {\bf (a)} ease of use, {\bf (b)} system trustworthiness, {\bf (c)} amount willing to pay, and {\bf (d)} whether the system made them more conscious of their cardiac health.}
  \label{fig:usability}
  \Description{This figure presents a series of four bar charts, each summarizing user perceptions from a survey of 23 participants. The charts are labeled (a) through (d) and are arranged horizontally. Chart (a), "Ease of use," shows a histogram with a y-axis representing "Count" from 0 to 10 and an x-axis representing a rating from 1 to 5. The bars are heavily skewed toward the higher end, with the majority of users rating the system at 5. Chart (b), "System trustworthiness," is similarly a histogram on a 1 to 5 scale. It also shows a positive distribution, with most users giving a rating of 4 or 5. Chart (c), "Amount willing to pay (USD)," shows a distribution of user willingness to pay. The y-axis shows "Count" up to 6, and the x-axis shows dollar amounts in bins of 0, 50, and 100 USD. The most frequent response is 0 USD, followed by 50 USD, and then 100 USD. Finally, chart (d), "More conscious of cardiac health," is a histogram with categorical responses: "Maybe," "Yes," and "No." The "Yes" category has the highest count, indicating that most users felt the system made them more conscious of their cardiac health.}
\end{figure*}

\noindent We evaluated user experience of our system through a short survey. \rr{23 of the 25 participants from the human subjects study consented to a post-study user experience survey that was sent over email, the remaining 2 participants did not respond. Informed written consent was obtained before the survey began.} Participants were sent a form containing the following questions:

\begin{enumerate}
\item Is the system easy to use? (1 = Not at all, 5 = Very easy)
\item Can you trust our system's heart sounds results? (1 = Not at all, 5 = Completely)
\item How much would you pay for this system? (USD)
\item Would this system make you more conscious of your cardiac health?
\item In what scenarios would you envision using this system?
\end{enumerate}

Fig.~\ref{fig:usability} summarizes the results of the survey. Participants rated the system as easy to use (mean \red{4.6 ± 0.6}) and  trustworthy (mean \red{4.3 ± 0.9}). The average amount participants would be willing to pay for the system was  \red{57.7 ± 58.9}, though this value likely reflects the socioeconomic status of our recruitment population, which primarily included individuals from our institution and surrounding community. In terms of whether the system would make one more conscious of cardiac health, \red{70\% ($n=16$)} said yes, \red{22\% ($n=5$)} said maybe, and \red{9\% ($n=2$)} said no.

Participants envisioned a wide range of scenarios for potential use, with two recurring themes from these results:
\textit{First}, participants were interested in the system's value to provide continuous monitoring during daily life and possibly even mitigate white coat effects of elevated cardiac measurements that can be found during clinical monitoring. One participant noted its possible value for cardiac disease prediction broadly beyond SCG and GCG monitoring given its potential to detect emergent conditions like cardiac arrest: \textit{When I am very old and alone, it could call an ambulance automatically if my heart stops.} We note that related emergency alert systems are integrated into commercial products, including smartwatches from Apple~\cite{fall_detection}, Samsung~\cite{fall_detection2}, and Google~\cite{fall_detection3}, which can automatically contact emergency services in the event of a fall or car crash.

\textit{Second}, participants were interested in the system's potential to monitor the effects of aerobic exercise, such as running, on overall cardiac activity. In the context of micro-cardiac monitoring, prior work~\cite{salerno1992exercise,korzeniowska2005usefulness} has noted that comparing SCG signals before and after exercise can useful for detecting coronary artery disease.




\section{Discussion}

\noindent {\bf Frequency and ease of recalibration.} The frequency at which the system needs to be recalibrated for an individual depends on changes in body state and skin properties over time which affect the body propagation channel. Such recalibration is common in both clinic-grade cardiac monitoring devices~\cite{jones2003measuring} and wrist-worn smart devices for cardiac monitoring from Samsung~\cite{swatch} and Aktiia~\cite{vybornova2021blood}, the latter two of which recommend monthly recalibration.

Our calibration procedure to account for physiological variation is lightweight: it only requires the user to wear their hearable and place their phone against their chest for approximately four seconds. Synchronization between the two devices can be achieved via an inaudible chirp transmitted from the phone and received by both devices, which is technique used for standardizing across sensor streams~\cite{wang2018seismo}.

\noindent \red{{\bf Privacy considerations.} While our system is designed for continuous operation, it is privacy-preserving by design and does not capture intelligible human speech. This is because human speech occupies frequencies > 80 Hz, with the majority of speech content between 300 and 3400 Hz~\cite{baken1987clinical}, which our system removes in software and can also suppress through an equivalent analog filter in hardware. Specifically, our design uses a bandpass filter to isolate ear-based cardiac signals < 50 Hz~\cite{rai2021comprehensive}. Furthermore, because hearables sit in the ear canal with rubber eartips or ear cups, they provide passive acoustic isolation that attenuates ambient sounds. Finally, all audio data are processed entirely on device in real time, and no data leave the user’s local devices.}

\noindent \red{{\bf Battery life.} We measured the battery consumption of our neural network model running in realtime on a Samsung Galaxy S9 smartphone using the ONNX runtime framework with NNAPI hardware acceleration to use the device's neural processing unit. When the model processes 800~ms chunks of audio data continuously, the battery drains at a rate of 8.2\% an hour, which equates to over 12 hours of uninterrupted monitoring. In comparison, when the model is not running and the phone remains idle, the battery drains at a rate of 7.6\% an hour. This shows that our inference pipeline only introduces a small overhead and is practical for continuous use.}

\noindent \red{{\bf Extended hearable use.} We note that our system does not require wearing hearables for extended durations to have health benefits. As it performs brief, opportunistic measurements during everyday moments when individuals wear hearables, such as commuting, it can provide a richer picture of cardiac health than episodic clinic visits.}

\subsection{\rr{Positioning and future directions in HCI}}

\rr{We position {\sysname} as a HCI contribution at the intersection of earable physiological sensing and personal informatics.}

\ssquishlist
\item \rr{{\bf Cardiac sensing with hearables.}  While there is rich prior HCI work that has explored cardiac monitoring from the ear~\cite{islam2025ballistobud,liu2025earmeter,christofferson2025artearial}, our key sensing approach is novel in its use of the {\it built-in speaker} which uniquely enables generalization across a broader range of devices, given that it is the only transducer common to all hearables. Despite speakers producing lower amplitude signals than microphones, we are able to achieve comparable results between the two transducers (Fig.~\ref{fig:cross_device_main}).} \rr{{\it{This distinction motivates the need for carefully designed interfaces that take into account the known attitudes, behavior, and psychology of users of mobile health technology~\cite{gabriele2020understanding} when designing trustworthy user interfaces.}}}

\rr{Micro mechanical cardiac monitoring with SCG and GCG remains under-explored in earables, and to the best of our knowledge no prior earable work has focused on constructing the timing of fiducial events corresponding to valve opening and closing. This focus matters because SCG and GCG encode cardiac mechanical timing, which complements prior work measuring electrical, acoustic and volumetric signals~\cite{cao2024earsteth,butkow2023heart,fan2023apg,li2024ecg,chen2024exploring,fan2021headfi,gilliam2022ear,liu2025earmeter,christofferson2025artearial}. While many cardiac signals are coupled through shared cardiovascular hemodynamics, they differ in physical origin, conventional sensing modality.}

\noindent \item \rr{{\bf Personal informatics.} A popular area in the HCI literature that our work can be categorized under is personal informatics. Work in this space has examined why people abandon self tracking tools and has produced multi-stage models of how tracking unfolds over time~\cite{epstein2016beyond,li2010stage,li2011understanding,choe2014designing}, including iterative stages spanning ``preparation, collection, integration, reflection, and action'', often noting that whether users are ``goal-oriented'' versus ``curiosity-oriented'' can affect how they approach these stages and how much they care about the underlying accuracy of the system, and the amount of data that is presented to them~\cite{rooksby2014personal}.}

\rr{Across this literature, a consistent theme is that self tracking breaks down when it depends on careful or repeated manual setup, when routines are disrupted, or when use becomes intermittent due to friction, device replacement, or unmet information needs~\cite{epstein2016beyond}.}

\rr{\textit{We view the micro-mechanical cardiac monitoring system through  a personal informatics lens, and not just as a sensing lens.} Work by Epstein et al.~\cite{epstein2016beyond,epstein2015lived} underscores the importance of developing systems that function “in the wild reality” with context switching, interruptions and inconsistency which are central to everyday self tracking. They further note that designing for these realities is central to a HCI contributions.}

\rr{Here, we design techniques that take into account different real-world issues that target specific breakdown points identified in the HCI personal informatics literature that prevent users from reaching ``reflection'' and ``action'' stages which are critical for behavior change. We show in Table~\ref{tab:phi_failure_modes} various such breakdowns and how we mitigate them.} \rr{\it{CHI learns that viewing physiological sensing through a personal informatics lens translates to design implications from the start to mitigate interruptions, device turnover, and trust breakdowns, which can pave the way for adoption.}}

\begin{table}[t]
\centering

{\color{black}
\begin{tabular}{p{0.42\linewidth} p{0.54\linewidth}}
\hline
\textbf{Common failure modes in personal health informatics} & \textbf{Design response} \\
\hline
Erosion of trust due to inconsistent readings &
Characterization of increased variability over conventional chest based IMU methods \\
\hline
Context switching and competing activities &
Evaluation during music playback; evaluation while lying down; motion artifact detection to enable automatic signal quality screening \\
\hline
Lapses in data quality after interruptions &
Ear worn placement supports repeatable remounting with minimal variability \\
\hline
Device replacement and turnover &
Zero effort cross device calibration enables continuity when the hearable is replaced or upgraded \\
\hline
Longitudinal drift and reduced comparability over time &
Stability evaluated across time; design of a periodic lightweight human in the loop calibration to maintain longitudinal comparability \\
\hline
\end{tabular}
}
\caption{\rr{Mapping common failure modes in personal health informatics systems identified by Epstein et al.~\cite{epstein2015lived,epstein2016beyond} to corresponding design responses in our system.}}
\label{tab:phi_failure_modes}
\end{table}

\rr{\subsubsection{Future directions in HCI} Future directions within the HCI community fall under three categories:}

\item \rr{{\bf Self-tracking and behavior change.} TummyTrials~\cite{karkar2017tummytrials} observed that those who attempt diagnostic self-tracking often lack knowledge to identify triggers that cause negative health conditions and thus may not reach their goals of managing their health conditions. Future iterations of our work can enable diagnostic self-tracking~\cite{rooksby2014personal} by mitigating common pitfalls identified from a community of self trackers~\cite{choe2014understanding} through two approaches:}

\rr{(1) We can implement lightweight activity recognition classifiers~\cite{montanari2023earset} on the recorded audio to capture the user’s context and behavior, instead of simply discarding it as a motion event such as in the current system implementation. This can help to draw correlations between changes in cardiac rhythm and the user’s activity to identify triggers that could worsen cardiac conditions such as the trigger of arrhythmias. \textit{Concretely, this shifts the system from producing isolated measurements to producing context-tagged episodes that better support ``reflection" and ``action" in personal informatics terms.}
}

\rr{This also aligns with the methodology Drunk User Interfaces~\cite{mariakakis2018drunk} which focused around the need to control for confounding effects such as user fatigue which would affect measurements of the endpoint of interest. In our case measuring pertinent variables such as posture, location, time of day, and producing a stratified performance reports to the user can help to ameliorate experimental confounds.}

\rr{(2) We can further build-upon self-experimentation frameworks~\cite{karkar2016framework} that can identify causal relationships between potential triggers and cardiac abnormalities through single-case experimental designs that generate a randomized study protocol that dictate the user’s level of physical activity on a given day.}

\item \rr{{\bf Clinical workflow integration.} A recurring theme in HCI research~\cite{west2018common} on patient generated data is that the value of self tracking does not automatically translate to value in clinical care. This is because data must cross boundaries between patient and clinician who view data with a different lens. Prior work~\cite{west2018common} shows that breakdowns often occur when patient and clinician goals are misaligned.}

\rr{HCI literature has framed interpretation of patient generated data as a collaborative sensemaking problem, emphasizing designs that reduce translation burden and support shared interpretation rather than expecting either party to independently derive meaning from raw data~\cite{raj2017understanding,chung2016boundary,islind2019shift,west2018common,pichon2021divided}.}

\rr{Building on this literature, an important future direction is to move beyond reliable sensing and explicitly support clinician mediated interpretation through interfaces that are aware of clinician workflows. We systematically provide future considerations and recommendations in Table~\ref{tab:west_barriers_design} using the barriers identified by West et al.~\cite{west2018common}.}

\begin{table}[t]
\centering
{\color{black}
\begin{tabular}{p{0.22\linewidth} p{0.72\linewidth}}
\hline
\textbf{Barrier} & \textbf{Implication for interfaces and outputs} \\
\hline
Structure and relevance &
Generate visit ready summaries using familiar clinical conventions. \\
\hline
Reliability and completeness &
Attach quality and coverage metadata that indicates when data was corrupted due to motion or low signal quality.\\
\hline
Context &
Future work here could include an activity recognition module to correlate cardiac events with activities performed by the user as well as user posture.\\
\hline
Time burden, overload, and interoperability &
Provide clinician dashboards with ``triage views" that compress high volume time series into a small set of markers of interest, and support export formats that integrate with documentation workflows. \\
\hline
\end{tabular}
}
\caption{\rr{Design implications aligned with barriers for patient generated data use in clinical care identified by West et al.~\cite{west2018common}.}}
\label{tab:west_barriers_design}
\end{table}

\item \rr{{\bf Human-AI partnerships.} Recent work has centered around the use of human-AI partnerships~\cite{van2025selective} to inform personal health decisions by acting as collaborative peers that could proactively and automatically surface health issues. In the context of our system this could be accomplished through automated interpretation of continuous time series data~\cite{englhardt2024classification} and integration with other digital phenotypes~\cite{insel2017digital} to extract behavioral patterns to surface results for review by users or clinical professionals.}

\rr{This is an exciting future direction with unique challenges at the technical and user-centered design perspective. For instance, literature has shown that there is a danger to being overly trusting of AI decisions compared to those mediated by medical professionals~\cite{guzman2020artificial}. Future work could focus on how such systems should incorporate known information about the user’s decision-making styles~\cite{scott1995decision}, the user’s health knowledge~\cite{dikmen2022effects}, pre-existing views about trust in AI~\cite{jacovi2021formalizing} to design a collaborative experience between users and clinicians.}

\ssquishend


\section{Limitations}

\noindent {\bf Open-ear earbuds.} We note that it is challenging to reliably record ear-based cardiac sounds from air-conduction open-ear earbuds~\cite{open_ear}. All of the hearables evaluated in our study provide some form of isolation against noise, either through a rubber ear tip seal, a passive ear cup, or a bone-conduction transducer. While open-ear devices comprise only a modest share of the hearables market, estimated at about 2\%~\cite{open_ear_market}, future designs could address these limitations by integrating active noise cancellation that compensate for the lack of passive isolation.


\noindent \red{{\bf Usage during motion.} Our cardiac sensing system is currently designed to operate during periods of stationary activity including sitting, standing, or lying down (e.g. during rest or sleep), and to ignore periods of motion, which overwhelm the cardiac signals measured at the ear. However, our system can still extract cardiac signals during brief pauses between movements such as short breaks during exercise, which capture the effect of exercise intensity on cardiac dynamics. We note while that motion artifact removal algorithms have been developed to extract heart rate from in-ear cardiac sounds using U-Net based architectures~\cite{butkow2023heart}, future work would be required to evaluate whether such techniques can be robustly applied to measure the micro-mechanical cardiac signals and the timing information of the fiducial points.}

\section{Future Directions}

\noindent {\bf Evaluation on subjects with cardiovascular conditions.} We present a proof-of-concept study to evaluate how SCG and GCG measurements could be captured across different hearable devices and users. Future studies are needed to perform a clinical validation of this technology in patients with or at risk of cardiovascular disease. These studies would be needed to assess if this technology can be used for the detection or diagnosis of cardiac conditions. These studies could build on prior clinical protocols~\cite{salerno1992exercise} that measure SCG before and after exercise to detect changes in cardiac function that could be used to detect coronary artery disease.

\noindent {\bf Joint reconstruction of micro-cardiac waveforms.} Although the SCG and GCG signals originate from the same underlying cardiac mechanics, they capture different motion components. In practice, one modality, often the GCG can have a lower SNR because rotational motion can be more subtle and more easily masked by sensor noise. Future work could explore a joint reconstruction framework that exploits complementary information between the two signals and leverages the stronger modality to guide reconstruction of the weaker signal. Prior work~\cite{yang2016annotation} has shown that gyroscopic readings can improve automatic annotation of SCG fiducial points and used to better estimate heart rate variability and cardiac time intervals.

\noindent {\bf Hearing aids.} Deployment of our system on hearing aids poses an interesting opportunity, particularly because these devices are widely adopted by older adults~\cite{chien2012prevalence} which is the same demographic at elevated risk for both cardiac disease and hearing loss~\cite{baiduc2023relationship,wattamwar2018association}. 

Traditional hearing aids such as those from Oticon~\cite{oticon} place in-ear speakers deep in the ear canal for sound amplification. In principle, these speakers would be able to capture the ear-based cardiac sounds robustly. The main challenge however, is that the hearing aid speakers continuously amplify ambient speech and sounds, and conventional speakers cannot simultaneously transmit and receive audio at high fidelity.

More recent devices such as the Apple AirPods Pro, which is licensed as a clinical-grade hearing aid~\cite{apple_hearing_aid}, incorporate both an in-ear microphone and speaker that would in principle be able to perform duplex operation. Here, the microphone could capture the cardiac sounds, while the speaker amplifies surrounding sounds. Given our evaluation (Sec.~\ref{sec:benchmark}) showing that the system works reliably even during music playback due to the non-overlapping frequencies of music and the cardiac sounds, having the system work on the Apple AirPods Pro remains a potentially exciting area of future work.

\bibliographystyle{ACM-Reference-Format}
\bibliography{mybib}

\end{document}